\documentclass[11pt]{article}

\usepackage{comment}
\usepackage[margin=1in]{geometry}
\usepackage{amsmath,amssymb,amsfonts,amsthm,mathtools}
\usepackage{algorithm}
\usepackage{algpseudocode}
\usepackage{natbib}

\mathtoolsset{showonlyrefs=true}
\numberwithin{equation}{section}
\numberwithin{algorithm}{section}
\allowdisplaybreaks

\newtheorem{proposition}{Proposition}[section]
\newtheorem{theorem}[proposition]{Theorem}
\newtheorem{lemma}[proposition]{Lemma}
\theoremstyle{definition}
\newtheorem{definition}[proposition]{Definition}
\theoremstyle{remark}
\newtheorem{remark}[proposition]{Remark}

\DeclareMathOperator*{\argmin}{arg\,min}
\DeclareMathOperator*{\argmax}{arg\,max}
\newcommand{\md}{\mathrm{d}}
\newcommand{\mE}{\mathbb{E}}
\newcommand{\mR}{\mathbb{R}}
\newcommand{\cH}{\mathcal{H}}
\newcommand{\cP}{\mathcal{P}}
\newcommand{\N}{\mathcal{N}}

\title{Optimal portfolio and consumption under no-shorting constraints:\\ The exploratory formulation and reinforcement learning}
\author{%
  Xun Li\thanks{Department of Applied Mathematics, The Hong Kong Polytechnic
  University. li.xun@polyu.edu.hk}
  \qquad Xiang Yu\thanks{Department of Applied Mathematics, The Hong Kong
  Polytechnic University. xiang.yu@polyu.edu.hk}
  \qquad Fengyi Yuan\thanks{School of Science and Engineering, the Chinese
  University of Hong Kong (Shenzhen). yuanfengyi@cuhk.edu.cn}
  \qquad Yijie Zeng\thanks{Department of Applied Mathematics, The Hong Kong
  Polytechnic University. yi-jie.zeng@connect.polyu.hk}
}
\date{}

\begin{document}
\maketitle

\begin{abstract}
We develop a reinforcement-learning framework for continuous-time optimal portfolio and consumption choice via utility maximization under no-shorting portfolio constraints. We adopt an entropy-regularized exploratory formulation and enforce the no-shorting requirement through a
componentwise nonnegative constraint on the conditional mean of admissible randomized portfolio policies, while the sampled actions are allowed to take negative values during
training.  For CRRA utility, we derive the homothetic value function and the
optimal Gaussian portfolio and Gibbs consumption policies under some model conditions and establish a verification theorem.  These analytical solutions motivate our parameterizations of the actor, the critic, and the
$q$-function for constrained policy iteration.  Policy evaluation uses
trajectory-based moment conditions, and policy improvement uses fitted
coefficients to solve a convex quadratic program over the nonnegative orthant
without requiring market coefficients as inputs.  The mean constraint
preserves tractable Gaussian exploration throughout parameter learning, while
modal deployment converts the learned randomized policies into deterministic
controls.  In particular, the Gaussian mode coincides with its constrained mean and hence
satisfies the no-shorting constraint, and the Gibbs mode determines the deployed consumption
rule.
\end{abstract}


\section{Introduction}

Continuous-time utility maximization for optimal portfolio and consumption choice has long been a central topic in mathematical finance. Within this literature, no-shorting portfolio constraints have attracted considerable attention due to the practical risk and regulation concerns. Empirical studies on the impact of no-shorting constraints on financial stability include \citet{Allen93}, \citet{Gomez}, \citet{Beber}, and \citet{Grullon}, among others. Theoretical developments in portfolio optimization under no-shorting constraints can be found in \citet{XS1992}, \cite{Shirakawa}, \cite{Cvit}, \cite{WYY}, and references therein.

Notably, conventional methods rely on accurate knowledge of the market parameters. When expected returns and covariances are unknown,
estimation errors may impair the out-of-sample performance of plug-in implementations of theoretically optimal controls. Reinforcement learning (RL) has emerged as a powerful paradigm that allows an agent to learn optimal dynamic decisions through direct interaction with an unknown environment. Fast growth has been seen in both RL theories and applications. To better account for some financial applications that evolve continuously along the timeline, increasing attention has been devoted to continuous-time RL. \citet{WangZariphopoulouZhou2020} pioneered a tractable exploratory formulation in continuous-time diffusion models via entropy regularization and provided a Gaussian distribution characterization of the optimal randomized policy. Later, \citet{jia2022policy}, \citet{JiaZhou2022PE}, \citet{JiaZhou2023} laid the foundations for continuous-time policy evaluation, policy gradient and q-learning algorithms. Recently, more theoretical extensions and applications of exploratory continuous-time RL have been considered in stochastic control problems, see, among others,
\citet{GXZ22}, \citet{Dong2023}, \citet{szpruch2024optimal}, \citet{wu2024reinforcement}, \citet{GRZ}, \citet{WYY24}, \citet{RL-Merton}, \citet{BHY25}, \citet{WY25}, \citet{FGLS}, \citet{WGL26}, \citet{GLZ26}, \citet{BHYZ26}.

The no-shorting policy constraint creates a new tension in this learning framework.  Full-support policy
randomization facilitates exploration, whereas a samplewise support constraint
changes the randomized policy class and can undermine analytical tractability. This
motivates a natural question: how can we preserve effective exploration, learn directly from
wealth--action data, and still obtain a deterministic no-shorting control for
deployment?

The primary goal of this paper is to address this challenge in the context of finite-horizon utility maximization for portfolio and consumption management. We consider the market model consisting of multiple correlated risky assets with unknown constant market coefficients. We impose no-shorting through a componentwise
nonnegative conditional-mean constraint on the randomized portfolio policy and first solve
the full-information exploratory problem under CRRA and logarithmic
utility. We then show the connection between the optimal randomized policy and the solution to the classical constrained Merton problem, which allows us to develop a constrained martingale
policy-iteration scheme.  The proposed algorithm evaluates policies from
observed trajectories, updates the portfolio through a nonnegative-orthant
quadratic program, and deploys the modes of the learned stochastic policies. 

Entropy regularization is the first ingredient of this construction. In our exploratory utility maximization problem, pointwise
maximization of the entropy-regularized Hamiltonian intrinsically produces a
Gibbs density whose exponent is determined by the local reward and the value
derivatives.  The quadratic portfolio Hamiltonian yields a Gaussian policy,
and the consumption Hamiltonian yields a Gibbs policy under CRRA utility and
a Gamma policy under logarithmic utility. These explicit forms give analytical parameterization of actor, critic, and little-$q$ function in learning algorithm.

The second ingredient is the nonnegative constraint on the conditional mean of the portfolio policy. In a related utility maximization setting, \citet{ChauNguyenNguyen2026} impose an
interval support constraint on a single-asset policy and obtain a truncated Gaussian distribution.
In a correlated multi-asset setting, however, orthant truncation yields a normalizing constant that generally does not factorize,
requiring multivariate Gaussian probability evaluations and more intricate sampling. Instead, by constraining only the conditional
mean, our formulation preserves a full multivariate Gaussian optimal policy. Its mean solves a covariance-coupled convex quadratic
program over the nonnegative orthant, and the required portfolio-action moments
stay explicit.  The quadratic program resolves binding constraints jointly
across correlated assets rather than clipping an unconstrained portfolio. Recently, 
\citet{LiWangZhou2026} considered the same policy-mean constraint in an unregularized
auxiliary problem for mean--variance selection with a time-varying and unknown
investment opportunity set, where the Gaussian covariance is specified
externally.  In our model, entropy
regularization determines the covariance jointly with the value function, and
the mean constraint reappears in policy improvement through the
nonnegative-orthant quadratic program.
%Because the randomized portfolio policy draws retain unrestricted pointwise support,
%the positive-temperature randomized problem remains distinct from the
%classical pointwise-constrained problem.

The third ingredient is modal deployment.  \citet{RL-Merton} consider the sampling of Gaussian
policies during training and execute the learned deterministic means, providing
the training--execution distinction that motivates our algorithm.  We deploy the
unique modes of both learned policy densities.  The Gaussian portfolio mode
equals its constrained mean and therefore satisfies the no-shorting constraint, although
individual training actions may have negative components.  As a key observation in this work, we reveal that when the model is known, the optimal portfolio mode recovers the classical
no-shorting portfolio at every positive temperature under logarithmic utility
and whenever the CRRA exploratory solution is well posed.  The Gamma mode also
recovers the classical logarithmic consumption rule.  For CRRA consumption,
the Gibbs mode is obtained directly from the fitted value coefficient, whereas
the policy mean generally has no closed form. Not only can we connect the optimal randomized policy under mean constraint to the classical optimal control via the mode, 
but we also bridge these two formulations via the convergence analysis when the temperature parameter tends to zero. \citet{TangZhangZhou2022} proved the vanishing entropy convergence from exploratory formulation to the
classical stochastic control problem  in the absence of constraint.  For the constrained utility maximization  in this work, we also obtain
exact positive-temperature modal identities and establish the
analogous zero-temperature convergence results for both exploratory value functions
and randomized policies.

The final ingredient is the martingale policy evaluation and policy iteration convergence under the portfolio-mean constraint tailored for our algorithm.  Building on the martingale policy-evaluation framework in
\citet{JiaZhou2022PE}, the $q$-learning in
\citet{JiaZhou2023}, and the grid-sampled policy construction in
\citet{JiaOuyangZhang2025}, we jointly characterize the value function and $q$-function through conditions that hold for every finite sampling grid.  This
characterization offers estimable moment equations based on observed
wealth--action trajectories, without relying on the market coefficients.  The fitted linear and quadratic $q$ function coefficients
determine a Gaussian portfolio update whose mean solves the
nonnegative-orthant quadratic program, while the consumption update retains
the Gibbs or Gamma form.  Furthermore, when the exact value function and $q$ function are used, we first prove that the constrained iteration improves the current
policy and then verify that the proposed iterations converge to the desired optimal policies once we start from some proper initial policies. Then, in the implementation step, replacing the exact q-function by its parameterized form allows the data-driven learning with satisfactory convergence performance on parameters.  

The above contributions are summarized as follows.
\begin{enumerate}
\item[(i)] In the exploratory utility maximization problem under a nonnegative mean constraint on portfolio policies, we derive and verify the optimal Gaussian portfolio policy and Gibbs or
Gamma consumption policy as well as some explicit CRRA well-posedness conditions and analytical expressions in the logarithmic case.

\item[(ii)] We establish exact modal recovery of the classical optimal portfolio and consumption under no-shorting constraints, which is new to the literature. Moreover, we establish the vanishing entropy convergence of the exploratory value functions and randomized policies under the nonnegative mean constraint on portfolio policies.

\item[(iii)] To provide the theoretical foundation for the algorithm design, we establish the theoretical policy iteration convergence in the constrained context when the model is known, in which the one-step iteration of portfolio policy solves a quadratic program. We also devise a model-free martingale policy evaluation step. When the exact q-function is replaced by its parameterization, the policy evaluation and policy iteration constitute our model-free RL algorithm.

\item[(iv)] We illustrate the $q$-function coefficient recovery, active-constraint
identification, and modal deployment in the implementation of our learning algorithm in two numerical experiments with satisfactory learning performance.
\end{enumerate}

The rest of the paper is organized as follows.  Section~\ref{sec:model}
introduces the market and auxiliary randomized controls.
Sections~\ref{sec:crra} and~\ref{sec:log} derive the optimal randomized
policies under CRRA and logarithmic utility, respectively.
Section~\ref{sec:modal-recovery} compares modal deployment with the classical problem.  Section~\ref{sec:policy-improvement} devises the
trajectory-based policy evaluation and constrained policy iteration convergence.  Section~\ref{sec:experiments} presents the numerical
experiments with satisfactory learning performance.  The appendices collect an auxiliary lemma on constrained entropy maximization and the proofs of the main results.  

\section{Market Model}\label{sec:model}

Fix a finite horizon $T>0$ and a filtered probability space
$(\Omega,\mathcal F,\mathbb F,\mathbb P)$ satisfying the usual conditions,
where $\mathbb F=(\mathcal F_t)_{0\le t\le T}$ supports an $m$-dimensional
Brownian motion $W$.  Consider a market model consisting of one risk-free asset with constant interest rate $r>0$ and $m$ risky assets whose price dynamics satisfy
\begin{equation}
  \frac{\md S_t^i}{S_t^i}
  =b_i\,\md t+\sum_{j=1}^m\sigma_{ij}\,\md W_t^j,
  \qquad i=1,\ldots,m.
\end{equation}
Here $b\in\mR^m$, and $\sigma\in\mR^{m\times m}$ is nonsingular.  Set
\begin{equation}
  e:=b-r\mathbf{1}_m,
  \qquad
  \rho:=\sigma^{-1}e,
  \qquad
  \Sigma:=\sigma\sigma^\top,
\end{equation}
where $\mathbf{1}_m$ is the all-ones vector.  The matrix $\Sigma$ is symmetric and
positive definite, and $\rho$ is the market price of risk.

At time $t\in[0,T]$, let $u\in\mR^m$ denote the vector of wealth
fractions invested in the risky assets and  let $c\in(0,\infty)$ be the
relative consumption rate.

For any density $f$ on its stated domain, let
\begin{equation}
  \cH(f):=-\int f(z)\log f(z)\,\md z
\end{equation}
denote its differential entropy whenever the integral is well defined, with
the convention $0\log0:=0$.
A randomized control is a pair of conditional densities
$\pi(\cdot\mid t,x)$ and $\eta(\cdot\mid t,x)$ for $u$ and $c$,
respectively, defined relative to Lebesgue measures $\md u$ and $\md c$.
We require the portfolio and consumption randomizations to be conditionally
independent given $(t,x)$.  Their joint density is therefore
$\pi(u\mid t,x)\eta(c\mid t,x)$, with entropy $\cH(\pi)+\cH(\eta)$.
We write
\begin{align}
  a_\pi(t,x)
  &:=\int_{\mR^m}u\,\pi(u\mid t,x)\,\md u,
  \\
  \bar c_\eta(t,x)
  &:=\int_0^\infty c\,\eta(c\mid t,x)\,\md c,\\
  M_\pi(t,x)
  &:=\int_{\mR^m}u^\top \Sigma u\,\pi(u\mid t,x)\,\md u.
\end{align}
The nonnegative-mean portfolio constraint in the present paper is imposed as
\begin{equation}
  a_\pi(t,x)\in\mR_+^m,
\end{equation}
where $\mR_+^m:=[0,\infty)^m$ denotes the nonnegative orthant.

For later use in the exploratory formulation, we also introduce the following classes of probability densities.
\begin{definition}\label{def:density-classes}
The portfolio density class $\cP_\pi(\mR^m)$ consists of the probability
densities $f$ on $\mR^m$ satisfying
\begin{equation}
  \int_{\mR^m}\lVert u\rVert^2f(u)\,\md u<\infty,
  \qquad
  \int_{\mR^m}u f(u)\,\md u\in\mR_+^m,
  \qquad
  \int_{\mR^m}f(u)|\log f(u)|\,\md u<\infty.
\end{equation}
The consumption density class $\cP_c$ consists of the probability densities
$f$ on $(0,\infty)$ satisfying
\begin{equation}
  \int_0^\infty c f(c)\,\md c<\infty,
  \qquad
  \int_0^\infty f(c)|\log f(c)|\,\md c<\infty.
\end{equation}
\end{definition}

We write $\N(a,Q)$ for the Gaussian law with mean $a$ and covariance $Q$,
identifying it with its density when convenient.

\begin{definition}\label{def:admissible-policy}
Fix a utility function $U\in C^1((0,\infty))$ and a positive Borel entropy
weight $\Lambda:[0,T]\times(0,\infty)\to(0,\infty)$.  A Markov randomized
policy pair $(\pi,\eta)$ is $(U,\Lambda)$-admissible if the following
conditions hold.
\begin{enumerate}
\item[(i)] The maps $(s,y,u)\mapsto\pi(u\mid s,y)$ and
$(s,y,c)\mapsto\eta(c\mid s,y)$ are jointly Borel measurable.  For every
$(s,y)\in[0,T]\times(0,\infty)$,
\begin{align}
  &\pi(\cdot\mid s,y)\in\cP_\pi(\mR^m),
  \qquad \eta(\cdot\mid s,y)\in\cP_c,\notag\\
  &\int_0^\infty |U(yc)|\eta(c\mid s,y)\,\md c<\infty.
\end{align}

\item[(ii)] For every $(t,x)\in[0,T]\times(0,\infty)$, there is a unique
law $\mathbb P_{t,x}^{\pi,\eta}$ on
$C([t,T];(0,\infty))$ under which the coordinate process $X^{\pi,\eta}$ starts from
$X_t^{\pi,\eta}=x$ and, for every $g\in C_c^2((0,\infty))$,
\begin{equation}
  g(X_s^{\pi,\eta})-g(x)-\int_t^s\mathcal L^{\pi,\eta}g(u,X_u^{\pi,\eta})\,\md u,
  \qquad s\in[t,T],
\end{equation}
is a local martingale, where
\begin{equation}
  \mathcal L^{\pi,\eta}g(s,y)
  :=y\left[r+e^\top a_\pi(s,y)-\bar c_\eta(s,y)\right]g'(y)
  +\frac{y^2}{2}M_\pi(s,y)g''(y).
\end{equation}
These laws form a Markov family and satisfy
\begin{equation}
  \int_t^T\left[\bar c_\eta(s,X_s^{\pi,\eta})+M_\pi(s,X_s^{\pi,\eta})\right]\,\md s<\infty,
  \qquad \mathbb P_{t,x}^{\pi,\eta}\text{-a.s.}
\end{equation}

\item[(iii)] Let $\mE_{t,x}^{\pi,\eta}$ be the expectation under
$\mathbb P_{t,x}^{\pi,\eta}$. We impose the integrability condition: 
\begin{align}
  \mE_{t,x}^{\pi,\eta}\Bigg[&|U(X_T^{\pi,\eta})|
  +\int_t^T\Bigg\{
  \int_0^\infty |U(X_s^{\pi,\eta}c)|\eta(c\mid s,X_s^{\pi,\eta})\,\md c \notag\\
  &\quad+\Lambda(s,X_s^{\pi,\eta})\left[
  |\cH(\pi(\cdot\mid s,X_s^{\pi,\eta}))|
  +|\cH(\eta(\cdot\mid s,X_s^{\pi,\eta}))|\right]
  \Bigg\}\,\md s\Bigg]<\infty.
\end{align}

\item[(iv)] The martingale admissibility condition
\begin{equation}
  \mE_{t,x}^{\pi,\eta}\left[
  \left(\int_t^T
  |X_s^{\pi,\eta}U'(X_s^{\pi,\eta})|^2M_\pi(s,X_s^{\pi,\eta})\,\md s\right)^{1/2}
  \right]<\infty
\end{equation}
holds for every $(t,x)\in[0,T]\times(0,\infty)$.
\end{enumerate}
For the CRRA case below, $U=U_\gamma$ and we choose
$\Lambda(s,y)=\lambda(s)y^{1-\gamma}$.  For the logarithmic case,
$U=\log$ and we choose $\Lambda(s,y)=\lambda$.  When the utility specification is
clear, we suppress this dependence and simply say admissible.
\end{definition}

The time-dependent coefficients of the CRRA and logarithmic verification
functions below are bounded on $[0,T]$.  Hence, Condition~(iv) and the
Burkholder--Davis--Gundy inequality ensure that their It\^o stochastic
integrals are uniformly integrable martingales.

\begin{remark}
The conditional-independence restriction entails no loss of optimality.  For
fixed portfolio and consumption marginals, the relaxed dynamics and utility
terms are unchanged, whereas the product coupling maximizes the joint entropy.
\end{remark}

Under an admissible policy pair $(\pi,\eta)$, the coordinate process in
Definition~\ref{def:admissible-policy} admits the following exploratory form:
\begin{equation}
  \begin{cases}
  \begin{aligned}
  \md X_s^{\pi,\eta}
  ={}&X_s^{\pi,\eta}
  \left[r+e^\top a_\pi(s,X_s^{\pi,\eta})
  -\bar c_\eta(s,X_s^{\pi,\eta})\right]\,\md s\\
  &+X_s^{\pi,\eta}
  M_\pi(s,X_s^{\pi,\eta})^{1/2}\,\md \widetilde W_s,
  \qquad s\in[t,T],
  \end{aligned}\\[0.4em]
  X_t^{\pi,\eta}=x,
  \end{cases}
  \label{eq:relaxed-wealth}
\end{equation}
where $\widetilde W$ is a one-dimensional Brownian motion.  

During training, randomized policies encourage exploration along sampled wealth trajectories. Once the policy parameters are learned, Proposition~\ref{prop:modal-recovery} allows us to deploy deterministic feedback portfolio and consumption controls by taking the modes of the learned densities.

To conclude this section, we provide a key quadratic program that will be crucial to our constrained RL framework:
\begin{equation}
  \widehat\mu
  :=\argmin_{\mu\in\mR_+^m}
  \left\{\frac12\mu^\top \Sigma\mu-e^\top\mu\right\}.
  \label{eq:mu-hat}
\end{equation}

\begin{proposition}\label{prop:qp-geometry}
Problem~\eqref{eq:mu-hat} has a unique solution, characterized by
\begin{equation}
  \widehat\mu\ge0,
  \qquad
  \ell:=\Sigma\widehat\mu-e\ge0,
  \qquad
  \ell^\top\widehat\mu=0.
\end{equation}
Moreover,
\begin{equation}
  \kappa
  :=\widehat\mu^\top\Sigma\widehat\mu
  =e^\top\widehat\mu
  \ge0.
\end{equation}
For every $a>0$,
\begin{align}
  \argmax_{u\in\mR_+^m}
  \left\{e^\top u-\frac{a}{2}u^\top\Sigma u\right\}
  &=\frac{\widehat\mu}{a},\label{eq:scailing1}\\
  \max_{u\in\mR_+^m}
  \left\{e^\top u-\frac{a}{2}u^\top\Sigma u\right\}
  &=\frac{\kappa}{2a}.\label{eq:scailing2}
\end{align}
\end{proposition}

\begin{proof}
See Appendix~\ref{app:proofs-model}.
\end{proof}

\section{Exploratory Formulation and Randomized Policies}
\label{sec:exploratory}

In this section, we introduce the entropy-regularized exploratory formulation, derive the exploratory HJB equation, and characterize the optimal randomized policies under CRRA and logarithmic utility. We also establish verification theorems.  These results suggest explicit parameterizations motivated by the analytical solutions.

\subsection{CRRA utility}\label{sec:crra}

Fix $\gamma>0$, $\gamma\ne1$, and let
\begin{equation}
  U_\gamma(x):=\frac{x^{1-\gamma}}{1-\gamma}.
\end{equation}
To preserve the homogeneity of the CRRA problem, we choose the state-dependent entropy parameter to be
\begin{equation}
  \Lambda(t,x):=\lambda(t)x^{1-\gamma},
  \qquad \lambda(t)>0.
\end{equation}
For an admissible policy pair $(\pi,\eta)$, define
\begin{align}
  J_\gamma^{\pi,\eta}(t,x)
  :=\mE_{t,x}\Bigg[&U_\gamma(X_T^{\pi,\eta})
  +\int_t^T\Bigg\{
  \int_0^\infty U_\gamma(X_s^{\pi,\eta}c)
  \eta(c\mid s,X_s^{\pi,\eta})\,\md c \notag\\
  &\quad+\Lambda(s,X_s^{\pi,\eta})
  \left[\cH(\pi(\cdot\mid s,X_s^{\pi,\eta}))
  +\cH(\eta(\cdot\mid s,X_s^{\pi,\eta}))\right]
  \Bigg\}\,\md s\Bigg].
  \label{eq:crra-objective}
\end{align}
The goal of the exploratory CRRA problem is to maximize the expected utility of terminal wealth and consumption, regularized by the entropy of the randomized policies.  Mathematically, the value function is
$V_\gamma(t,x):=\sup_{(\pi,\eta)}
J_\gamma^{\pi,\eta}(t,x)$, where the supremum is taken over the
$(U_\gamma,\Lambda)$-admissible Markov policy pairs of
Definition~\ref{def:admissible-policy}.

Suppressing the conditioning variables in the pointwise densities, the HJB
equation associated with \eqref{eq:crra-objective} separates into portfolio
and consumption maximizations. More specifically,
\begin{equation}
  \begin{cases}
  \begin{aligned}
  0={}&\partial_tV_\gamma+rx\partial_xV_\gamma \\
  &+\sup_{\pi\in\cP_\pi(\mR^m)}\Bigg\{
  \int_{\mR^m}\left[
  x\partial_xV_\gamma\,e^\top u
  +\frac{x^2\partial_{xx}V_\gamma}{2}u^\top \Sigma u
  \right]\pi(u)\,\md u
  +\Lambda(t,x)\cH(\pi)\Bigg\} \\
  &+\sup_{\eta\in\cP_c}\Bigg\{
  \int_0^\infty\left[
  U_\gamma(xc)-x\partial_xV_\gamma\,c
  \right]\eta(c)\,\md c
  +\Lambda(t,x)\cH(\eta)\Bigg\},
  \end{aligned} \\[0.6em]
  V_\gamma(T,x)=U_\gamma(x).
  \end{cases}
  \label{eq:crra-hjb}
\end{equation}
To facilitate the analysis of \eqref{eq:crra-hjb}, we provide several preliminary technical results.

For $h,\lambda>0$, introduce the partition function associated with the
consumption maximization in \eqref{eq:crra-hjb}:
\begin{equation}
  Z_\gamma(h,\lambda)
  :=\int_0^\infty
  \exp\left\{\frac{1}{\lambda}
  \left(\frac{c^{1-\gamma}}{1-\gamma}-hc\right)\right\}\,\md c.
\end{equation}

\begin{lemma}\label{lem:crra-partition}
For every $\gamma>0$, $\gamma\ne1$, the partition function satisfies
$0<Z_\gamma(h,\lambda)<\infty$ on $(0,\infty)^2$.  It is jointly continuous
in $(h,\lambda)$, and its partial derivative $\partial_hZ_\gamma$ exists and
is jointly continuous.  Define
\begin{equation}
  \eta_{h,\lambda}(c)
  :=\frac{1}{Z_\gamma(h,\lambda)}
  \exp\left\{\frac{1}{\lambda}
  \left(\frac{c^{1-\gamma}}{1-\gamma}-hc\right)\right\}
  \mathbf{1}_{(0,\infty)}(c).
\end{equation}
For every compact set $K\subset(0,\infty)^2$,
\begin{equation}
  \sup_{(h,\lambda)\in K}\int_0^\infty
  \left(c+c^{1-\gamma}+|\log\eta_{h,\lambda}(c)|\right)
  \eta_{h,\lambda}(c)\,\md c<\infty.
\end{equation}
\end{lemma}

\begin{proof}
See Appendix~\ref{app:proofs-exploratory}.
\end{proof}

For a continuous positive temperature schedule, write
\begin{equation}
  \lambda_-:=\min_{t\in[0,T]}\lambda(t),
  \qquad
  \lambda_+:=\max_{t\in[0,T]}\lambda(t),
  \label{eq:crra-temperature-range}
\end{equation}
and set
\begin{equation}
  \mathfrak G_{\gamma,\lambda}(t,h)
  :=\left(r+\frac{\kappa}{2\gamma}\right)h
  +\frac{\lambda(t)}{2}
  \log\left(
  \frac{(2\pi\lambda(t))^m}
  {(\gamma h)^m\det(\Sigma)}
  \right)
  +\lambda(t)\log Z_\gamma(h,\lambda(t)),
  \qquad h>0.
  \label{eq:crra-G-definition}
\end{equation}
By Lemma~\ref{lem:crra-partition},
\begin{equation}
  \partial_h\mathfrak G_{\gamma,\lambda}(t,h)
  =r+\frac{\kappa}{2\gamma}-\frac{m\lambda(t)}{2h}
  -\int_0^\infty c\,\eta_{h,\lambda(t)}(c)\,\md c.
\end{equation}
This derivative is continuous on $[0,T]\times(0,\infty)$ and bounded on
$[0,T]\times[a,b]$ for every $0<a<b<\infty$.  Hence
$\mathfrak G_{\gamma,\lambda}$ is continuous and locally Lipschitz in $h$,
locally uniformly in $t$.  The following auxiliary ODE will play a key role in analyzing the CRRA exploratory HJB equation \eqref{eq:crra-hjb}:
\begin{equation}
\begin{cases}
  \displaystyle
  \frac{h'(t)}{1-\gamma}
  +\mathfrak G_{\gamma,\lambda}(t,h(t))=0,\\
  h(T)=1.
\end{cases}
\label{eq:crra-h-ode}
\end{equation}

\begin{theorem}\label{thm:crra-ode}
Suppose that $\lambda\in C([0,T])$ and $\lambda(t)>0$ for every
$t\in[0,T]$.  There exists a
unique maximal $C^1$ solution to \eqref{eq:crra-h-ode}, with values in $(0,\infty)$, defined locally on a backward
interval near $T$. If $0<\gamma<1$, this solution extends to all of
$[0,T]$.  If $\gamma>1$ and there exists $\underline h\in(0,1]$ such that
\begin{equation}
  \mathfrak G_{\gamma,\lambda}(t,\underline h)\le0,
  \qquad t\in[0,T],
  \label{eq:crra-lower-barrier-condition}
\end{equation}
then the solution extends to all of $[0,T]$ and satisfies
$h(t)\ge\underline h$ for every $t\in[0,T]$.
\end{theorem}

\begin{proof}
See Appendix~\ref{app:proofs-exploratory}.
\end{proof}

The next result gives an explicit low-temperature sufficient condition for
\eqref{eq:crra-lower-barrier-condition} in terms of the model coefficients.

\begin{proposition}\label{prop:crra-lower-barrier}
Suppose that $\gamma>1$ and $\lambda\in C([0,T])$ is strictly positive.
Define
\begin{equation}
  \underline h
  :=\min\left\{1,
  \left[
  \frac{\gamma}
  {2(\gamma-1)\left(r+\kappa/(2\gamma)\right)}
  \right]^\gamma\right\}.
  \label{eq:crra-explicit-lower-barrier}
\end{equation}
Then $\underline h$ satisfies \eqref{eq:crra-lower-barrier-condition} whenever
\begin{align}
  \lambda_+\le\min\Bigg\{&
  \underline h^{(\gamma-1)/\gamma},\notag\\
  &\frac{\gamma\underline h^{(\gamma-1)/\gamma}}
  {2(\gamma-1)\left[
  \frac12\left|\log\left(
  \frac{(2\pi)^m}
  {\gamma^m\underline h^{m+2/\gamma}\det(\Sigma)}
  \right)\right|
  +\log\left(\frac{2\gamma-1}{\gamma-1}\right)
  \right]}
  \Bigg\}.
  \label{eq:crra-low-temperature-condition}
\end{align}
\end{proposition}

\begin{proof}
See Appendix~\ref{app:proofs-exploratory}.
\end{proof}

Together, Theorem~\ref{thm:crra-ode} and Proposition \ref{prop:crra-lower-barrier} show that requiring
a positive solution to \eqref{eq:crra-h-ode} imposes no additional structural
restriction on the model coefficients under the standing assumptions.
Positivity is trivial when $0<\gamma<1$, whereas, for $\gamma>1$, the threshold in
\eqref{eq:crra-low-temperature-condition} is strictly positive for every
fixed collection of coefficients, so it suffices that the regularization
temperature be uniformly small.  These results thus completely answer the question of well-posedness in a general formulation incorporating  randomized
consumption, multiple assets, a time-varying temperature, and a policy-mean
no-shorting constraint.

The following proposition establishes one of the key results of this paper, i.e., identifying the optimal randomized policies under entropy regularization and nonnegative-mean constraints.
\begin{proposition}\label{prop:crra-hjb-solution}
Suppose that a positive continuously differentiable function $h$ solves
\eqref{eq:crra-h-ode}.
Then the homothetic solution of \eqref{eq:crra-hjb} is
\begin{equation}
  V_\gamma(t,x)=\frac{h(t)}{1-\gamma}x^{1-\gamma},
  \label{eq:crra-value}
\end{equation}
and its pointwise maximizing randomized policies are
\begin{align}
  \pi_\gamma^*(\cdot\mid t,x)
  &=\N\left(
  \frac{1}{\gamma}\widehat\mu,
  \frac{\lambda(t)}{\gamma h(t)}\Sigma^{-1}
  \right),
  \label{eq:crra-portfolio}\\
  \eta_\gamma^*(c\mid t,x)
  &=\frac{1}{Z_\gamma(h(t),\lambda(t))}
  \exp\left\{\frac{1}{\lambda(t)}
  \left(\frac{c^{1-\gamma}}{1-\gamma}-h(t)c\right)\right\}
  \mathbf{1}_{(0,\infty)}(c).
  \label{eq:crra-consumption}
\end{align}
\end{proposition}

\begin{proof}
See Appendix~\ref{app:proofs-exploratory}.
\end{proof}

\begin{remark}
The logarithm of the density in \eqref{eq:crra-consumption} is strictly
concave and has derivative
$(c^{-\gamma}-h(t))/\lambda(t)$.  Hence its unique mode is
$m(t):=h(t)^{-1/\gamma}$, which gives the deployed relative consumption rate
and the consumption amount $xm(t)$ at wealth $x$.  One advantage of modal deployment is that it avoids an additional parameterization of the actor in the learning algorithm of Section~\ref{sec:policy-improvement}.
For example, the fitted critic coefficient $h^\theta(t)$ in
\eqref{eq:data-crra-critic} enters the Gibbs consumption update
\eqref{eq:data-crra-consumption-update} directly, yielding the deployed mode
$(h^\theta(t))^{-1/\gamma}$.
By contrast, the corresponding policy mean is a derivative of
$\log Z_\gamma$ and generally has no closed form, so a mean-parameterized
update would require numerical inversion to recover the mode.
Section~\ref{sec:modal-recovery} further shows that the optimal modal rate
$h_\varepsilon^{-1/\gamma}$ converges to the optimal rate of the original
model under zero-temperature annealing, providing an additional operational
motivation for modal deployment.
\end{remark}

We next verify the admissibility and optimality of the candidate randomized
policies in Proposition~\ref{prop:crra-hjb-solution}.  The key step is an
explicit entropy-gap identity, after which the usual It\^o verification
argument applies.
For densities $q_1$ and $q_2$, write
$\operatorname{KL}(q_1\Vert q_2)
:=\int q_1(z)\log(q_1(z)/q_2(z))\,\md z$ whenever it is well defined.

\begin{theorem}\label{thm:crra-verification}
Suppose that $\lambda\in C([0,T])$ is strictly positive and that
\eqref{eq:crra-h-ode} admits a positive $C^1$ solution $h$.  Then the policy
pair $(\pi_\gamma^*,\eta_\gamma^*)$ in \eqref{eq:crra-portfolio}--
\eqref{eq:crra-consumption} is admissible and optimal for
\eqref{eq:crra-objective}.  The value function is $V_\gamma$ in
\eqref{eq:crra-value}.
\end{theorem}

\begin{proof}
See Appendix~\ref{app:proofs-exploratory}.
\end{proof}

\subsection{Logarithmic utility}\label{sec:log}

For logarithmic utility, fix a constant regularization temperature
$\lambda>0$.  For an admissible policy pair $(\pi,\eta)$, define
\begin{align}
  J_{\log}^{\pi,\eta}(t,x)
  :=\mE_{t,x}\Bigg[&\log X_T^{\pi,\eta}
  +\int_t^T\Bigg\{
  \int_0^\infty\log(X_s^{\pi,\eta}c)
  \eta(c\mid s,X_s^{\pi,\eta})\,\md c\notag\\
  &\quad+\lambda\left[
  \cH(\pi(\cdot\mid s,X_s^{\pi,\eta}))
  +\cH(\eta(\cdot\mid s,X_s^{\pi,\eta}))
  \right]\Bigg\}\,\md s\Bigg].
  \label{eq:log-objective}
\end{align}
Let $V_{\log}$ denote the value function associated with
\eqref{eq:log-objective}, where the supremum is taken over the
$(\log,\lambda)$-admissible Markov policy pairs of
Definition~\ref{def:admissible-policy}.  As in the CRRA problem, the
corresponding HJB equation separates into portfolio and consumption
maximizations.  Here the identity $\log(xc)=\log x+\log c$ makes the state
dependence additive, which motivates the additive ansatz
\begin{equation}
  V_{\log}(t,x)=g(t)\log x+k(t),
  \qquad g(T)=1, \quad k(T)=0.
  \label{eq:log-ansatz}
\end{equation}
Substitution into the HJB equation and matching the coefficient of $\log x$
gives
\begin{equation}
  g(t)=1+T-t.
\end{equation}
The partition function for the consumption maximization is explicit:
\begin{equation}
  Z_{\log}(t)
  :=\int_0^\infty c^{1/\lambda}
  \exp\left(-\frac{g(t)}{\lambda}c\right)\,\md c
  =\Gamma\left(1+\frac{1}{\lambda}\right)
  \left(\frac{\lambda}{g(t)}\right)^{1+1/\lambda}.
\end{equation}

\begin{proposition}\label{prop:log-hjb-solution}
The unique pointwise maximizing portfolio and consumption policies are
\begin{align}
  \pi_{\log}^*(\cdot\mid t,x)
  &=\N\left(
  \widehat\mu,
  \frac{\lambda}{g(t)}\Sigma^{-1}
  \right),
  \label{eq:log-portfolio}\\
  \eta_{\log}^*(\cdot\mid t,x)
  &=\operatorname{Gamma}\left(
  1+\frac{1}{\lambda},\frac{g(t)}{\lambda}
  \right),
  \label{eq:log-consumption}
\end{align}
where the Gamma distribution is parameterized by shape and rate.  Equivalently,
\begin{equation}
  \eta_{\log}^*(c\mid t,x)
  =\frac{1}{Z_{\log}(t)}c^{1/\lambda}
  \exp\left(-\frac{g(t)}{\lambda}c\right)
  \mathbf{1}_{(0,\infty)}(c).
  \label{eq:log-consumption-density}
\end{equation}
The additive term in \eqref{eq:log-ansatz} is
\begin{equation}
  k(t)=\int_t^T\Bigg[g(s)\left(r+\frac{\kappa}{2}\right)
  +\frac{\lambda}{2}\log\left(
  \frac{(2\pi\lambda)^m}{g(s)^m\det(\Sigma)}
  \right)
  +\lambda\log Z_{\log}(s)\Bigg]\,\md s.
  \label{eq:log-k}
\end{equation}
\end{proposition}

\begin{proof}
See Appendix~\ref{app:proofs-exploratory}.
\end{proof}

We next verify that the pointwise maximizers in
Proposition~\ref{prop:log-hjb-solution} solve the exploratory control problem.

\begin{theorem}\label{thm:log-verification}
For every $\lambda>0$, the policy pair in
\eqref{eq:log-portfolio}--\eqref{eq:log-consumption} is admissible and optimal
for \eqref{eq:log-objective}.  The function in
\eqref{eq:log-ansatz}, with $g(t)=1+T-t$ and $k$ given by
\eqref{eq:log-k}, is the value function.
\end{theorem}

\begin{proof}
See Appendix~\ref{app:proofs-exploratory}.
\end{proof}

The explicit Gamma form also gives the modal execution rule.  The policy in
\eqref{eq:log-consumption} has the unique mode $1/g(t)$, which is the deployed
relative consumption rate.  Conditional on $(t,x)$, the auxiliary randomized
consumption amount $xc$ is Gamma distributed with shape $1+1/\lambda$ and rate
$g(t)/(\lambda x)$.  Its mode, and hence the deployed consumption amount, is
$x/g(t)=x/(1+T-t)$.

\section{Connection to the Classical Constrained Merton Problem}\label{sec:modal-recovery}

This section relates the exploratory formulation to the classical Merton problem with no-shorting portfolio constraint. The classical benchmark enforces the
no-shorting constraint pointwise on the portfolio control, whereas the exploratory
formulation allows action draws in $\mR^m$ and constrains only the conditional mean. We first recall the solution of the benchmark and then
relate it to the exploratory solution through exact mode identities. As both results are rather standard, we omit the proofs.

For progressively measurable controls
$u_s\in\mR_+^m$ and $c_s>0$, let
\begin{equation}
  \begin{cases}
  \begin{aligned}
  \md X_s^{u,c}
  ={}&X_s^{u,c}\left[r+e^\top u_s-c_s\right]\,\md s\\
  &+X_s^{u,c}u_s^\top\sigma\,\md W_s,
  \qquad s\in[t,T],
  \end{aligned}\\[0.4em]
  X_t^{u,c}=x.
  \end{cases}
\end{equation}
A control pair $(u,c)$ is admissible if this system has a nonexplosive,
strictly positive solution, the objective below is finite, and the localized
stochastic integrals used in verification are uniformly integrable.  For
either CRRA or logarithmic utility, define
\begin{equation}
  J_U^0(t,x;u,c)
  :=\mE_{t,x}\left[
  U(X_T^{u,c})+\int_t^T U(X_s^{u,c}c_s)\,\md s
  \right],
  \qquad
  V_U^0(t,x):=\sup_{u\ge0,\,c>0}J_U^0(t,x;u,c).
\end{equation}
Here the supremum is taken over admissible controls.  The superscript $0$
distinguishes this unregularized Merton benchmark from the positive-temperature
exploratory value functions.

\begin{proposition}\label{prop:deterministic-solution}
The classical constrained Merton benchmark has the following explicit
solution.  For CRRA utility, fix $\gamma>0$, $\gamma\ne1$, and set
\begin{align}
  \beta_\gamma
  &:=\frac{1-\gamma}{\gamma}
  \left(r+\frac{\kappa}{2\gamma}\right),\\
  h_0(t)
  &:=\left(
  e^{\beta_\gamma(T-t)}
  +\int_t^T e^{\beta_\gamma(s-t)}\,\md s
  \right)^\gamma.
\end{align}
Then $h_0$ is the unique $C^1$ solution with values in $(0,\infty)$ of
\begin{equation}
  \begin{cases}
  \displaystyle
  \frac{h_0'(t)}{1-\gamma}
  +\left(r+\frac{\kappa}{2\gamma}\right)h_0(t)
  +\frac{\gamma}{1-\gamma}
  h_0(t)^{(\gamma-1)/\gamma}=0,\\
  h_0(T)=1.
  \end{cases}
\end{equation}
The CRRA value function and optimal controls are
\begin{equation}
  V_\gamma^0(t,x)
  =\frac{h_0(t)}{1-\gamma}x^{1-\gamma},
  \qquad
  u_\gamma^0(t)=\frac{\widehat\mu}{\gamma},
  \qquad
  c_\gamma^0(t)=h_0(t)^{-1/\gamma}.
\end{equation}
For logarithmic utility, let $g(t)=1+T-t$.  The value function and optimal
controls are
\begin{align}
  V_{\log}^0(t,x)
  &=g(t)\log x+k_0(t),\\
  k_0(t)
  &=\int_t^T\left[
  g(s)\left(r+\frac{\kappa}{2}\right)-\log g(s)-1
  \right]\,\md s,\\
  u_{\log}^0(t)
  &=\widehat\mu,
  \qquad
  c_{\log}^0(t)=\frac{1}{g(t)}.
\end{align}
In both cases, the optimal controls are unique up to
Lebesgue--$\mathbb P$ product-null sets.
\end{proposition}

\begin{proposition}\label{prop:modal-recovery}
For the CRRA problem, suppose that $\lambda\in C([0,T])$ is strictly positive
and \eqref{eq:crra-h-ode} admits a positive $C^1$ solution $h$.  Then the
optimal randomized portfolio policy has the unique mode
\begin{equation}
  \argmax_{u\in\mR^m}\pi_\gamma^*(u\mid t,x)
  =u_\gamma^0(t),
  \qquad (t,x)\in[0,T]\times(0,\infty).
\end{equation}
For the logarithmic problem with any constant temperature $\lambda>0$, the
optimal randomized portfolio and consumption policies have the unique modes
\begin{align}
  \argmax_{u\in\mR^m}\pi_{\log}^*(u\mid t,x)
  &=u_{\log}^0(t),\\
  \argmax_{c>0}\eta_{\log}^*(c\mid t,x)
  &=c_{\log}^0(t),
  \qquad (t,x)\in[0,T]\times(0,\infty).
\end{align}
Thus, under the previous conditions, modal execution recovers the classical
optimal portfolio under both utility specifications.  Under logarithmic
utility, it recovers the full classical optimal control pair.
\end{proposition}

For CRRA consumption, the
mode $h(t)^{-1/\gamma}$ can also differ from the classical rate
$c_\gamma^0(t)=h_0(t)^{-1/\gamma}$.  The next result shows that this discrepancy
vanishes as the temperature tends to zero, establishes concentration of the
randomized policies on the classical optimal controls, and gives convergence
of both exploratory value functions to their unregularized counterparts.

\begin{theorem}\label{thm:zero-temperature}
Let $\bar\lambda\in C([0,T])$ be strictly positive and set
$\lambda_\varepsilon(t)=\varepsilon\bar\lambda(t)$.  All logarithmic objects
below are evaluated at the constant temperature $\lambda=\varepsilon$.
For every $\gamma>0$, $\gamma\ne1$, there exist
$\varepsilon_0\in(0,e^{-1})$ and $C>0$ such that the following statements
hold for $0<\varepsilon\le\varepsilon_0$, where $h_\varepsilon$ denotes the
positive solution of \eqref{eq:crra-h-ode} with temperature
$\lambda_\varepsilon$.
\begin{enumerate}
\item[(i)] \emph{CRRA value convergence.} We have
\begin{equation}
  \|h_\varepsilon-h_0\|_\infty
  \le C\varepsilon|\log\varepsilon|.
\end{equation}
Moreover, with $V_{\gamma,\varepsilon}(t,x):=\frac{h_\varepsilon(t)}{1-\gamma}x^{1-\gamma}$,
for every compact set $K\subset[0,T]\times(0,\infty)$, there exists
$C_K>0$, independent of $\varepsilon$, such that
\begin{equation}
  \sup_{(t,x)\in K}
  |V_{\gamma,\varepsilon}(t,x)-V_\gamma^0(t,x)|
  \le C_K\varepsilon|\log\varepsilon|.
\end{equation}

\item[(ii)] \emph{CRRA policy convergence.}
Let $\pi_{\gamma,\varepsilon}^*$ and
$\eta_{\gamma,\varepsilon}^*$ denote the policies in
\eqref{eq:crra-portfolio}--\eqref{eq:crra-consumption} evaluated at
$h=h_\varepsilon$ and $\lambda=\lambda_\varepsilon$.  The deployed
consumption mode satisfies
\begin{equation}
  \sup_{t\in[0,T]}
  \left|h_\varepsilon(t)^{-1/\gamma}
  -c_\gamma^0(t)\right|
  \le C\varepsilon|\log\varepsilon|.
\end{equation}
Moreover, for every $x>0$,
\begin{align}
  \sup_{t\in[0,T]}\int_{\mR^m}
  |u-u_\gamma^0(t)|\,
  \pi_{\gamma,\varepsilon}^*(u\mid t,x)\,\md u
  &\le C\sqrt{\varepsilon},\\
  \sup_{t\in[0,T]}\int_0^\infty
  |c-c_\gamma^0(t)|\,
  \eta_{\gamma,\varepsilon}^*(c\mid t,x)\,\md c
  &\le C\sqrt{\varepsilon}.
\end{align}

\item[(iii)] \emph{Logarithmic value convergence.}
Let $V_{\log,\varepsilon}$ be the value function in
\eqref{eq:log-ansatz}--\eqref{eq:log-k}.  Then
\begin{equation}
  \sup_{(t,x)\in[0,T]\times(0,\infty)}
  |V_{\log,\varepsilon}(t,x)-V_{\log}^0(t,x)|
  \le C\varepsilon|\log\varepsilon|.
\end{equation}

\item[(iv)] \emph{Logarithmic policy convergence.}
Let $\pi_{\log,\varepsilon}^*$ and $\eta_{\log,\varepsilon}^*$ denote the
policies in \eqref{eq:log-portfolio}--\eqref{eq:log-consumption}.  Then, for
every $x>0$,
\begin{align}
  \sup_{t\in[0,T]}\int_{\mR^m}
  |u-u_{\log}^0(t)|\,
  \pi_{\log,\varepsilon}^*(u\mid t,x)\,\md u
  &\le C\sqrt{\varepsilon},\\
  \sup_{t\in[0,T]}\int_0^\infty
  |c-c_{\log}^0(t)|\,
  \eta_{\log,\varepsilon}^*(c\mid t,x)\,\md c
  &\le C\sqrt{\varepsilon}.
\end{align}
\end{enumerate}
\end{theorem}

\begin{proof}
See Appendix~\ref{app:proofs-modal-recovery}.
\end{proof}

\begin{remark}
Proposition~\ref{prop:modal-recovery} and
Theorem~\ref{thm:zero-temperature} provide a comparison with the benchmark solution of the
classical Merton problem.  The former gives exact identities for deployed modes at
positive temperature.  The latter shows that the first absolute deviations
of the randomized actions from the classical controls vanish uniformly in
time and gives quantitative convergence of both value functions.
The partition functions of the portfolio and consumption policies contribute
$((m+1)/2)\lambda\log\lambda$ to the HJB equation, which produces the
$\varepsilon|\log\varepsilon|$ scale in value and mode bounds.  The randomized
policies, on the other hand, converge at the standard-deviation rate $O(\sqrt\varepsilon)$.
\end{remark}

\section{Policy Evaluation and Policy Iteration Convergence}
\label{sec:policy-improvement}

This section develops policy evaluation and policy improvement and convergence for the learning algorithm.  The evaluation step identifies the value
function and the \(q\)-function from a martingale condition, following
\citet{JiaZhou2023} and its official erratum.  The improvement iteration step converts
the identified \(q\) function coefficients into a no-shorting-constrained stochastic policy. The iteration convergence towards the optimal policy is established when the model is known. 

Throughout this section, \(U\) denotes either \(U_\gamma\) or \(\log\),
\(J_U^{\pi,\eta}\) denotes the corresponding fixed-policy objective in
\eqref{eq:crra-objective} or \eqref{eq:log-objective}, and \(\Lambda\) denotes
the associated entropy weight.

\subsection{Martingale policy evaluation}

For an exploratory action process \((A,C)\), let the observed wealth satisfy
\begin{equation}
  \md X_s=X_s\left[r+e^\top A_s-C_s\right]\md s
  +X_sA_s^\top\sigma\,\md W_s,\qquad X_t=x.
  \label{eq:sampled-wealth}
\end{equation}
Let \(X^{u,c}\) denote the wealth process starting from \(x\) at time \(t\)
when the action pair \((u,c)\) is held fixed on \([t,t+\Delta]\). Write
\(\mE_{t,x}^{u,c}\) for the corresponding expectation.  For an admissible pair
\((\pi,\eta)\), define its little-\(q\) function by
\begin{equation}
  q_U^{\pi,\eta}(t,x,u,c)
  :=\lim_{\Delta\downarrow0}\frac{1}{\Delta}
  \mE_{t,x}^{u,c}\left[
  J_U^{\pi,\eta}(t+\Delta,X_{t+\Delta}^{u,c})
  -J_U^{\pi,\eta}(t,x)
  +\int_t^{t+\Delta}U(X_s^{u,c}c)\,\md s
  \right],
  \label{eq:little-q-definition}
\end{equation}
whenever the limit exists.  The transition in
\eqref{eq:little-q-definition} is observed, so the algorithm does not need to evaluate
the actionwise generator.

Throughout this section, a policy pair is called scale invariant if
$\pi(u\mid t,x)=\pi_t(u)$ and $\eta(c\mid t,x)=\eta_t(c)$.  For such a pair,
write $\nu_t(u,c):=\pi_t(u)\eta_t(c)$ for its joint density.  It has full
support if \(\nu_t(u,c)>0\) for Lebesgue-a.e.
\((u,c)\in\mR^m\times(0,\infty)\) and every \(t<T\).

Fix a grid \(\mathcal G=\{t=t_0<\cdots<t_N=T\}\).  At \(t_i\), use an
independent auxiliary seed to draw \((A_i,C_i)\sim\nu_{t_i}\), conditionally
on current information and independently of future market noise, and hold the
action on \([t_i,t_{i+1})\).  Denote the resulting piecewise-constant actions
by \((A^{\mathcal G},C^{\mathcal G})\) and the solution of
\eqref{eq:sampled-wealth} by \(X^{\mathcal G}\), as in
\citet{JiaOuyangZhang2025}.  Writing \(\xi_i\) for the seed revealed at
\(t_i\), let \(\mathbb F^{\mathcal G}\) be the usual augmentation of
\((\mathcal F_s\vee\sigma(\xi_i:t_i\le s))_{t\le s\le T}\).  The seed
independence ensures that \(W\) remains a Brownian motion in this filtration,
and the sampled action process is adapted and piecewise constant.  Set
\begin{equation*}
  \mathcal V:=C^{1,2}([0,T)\times(0,\infty))
  \cap C([0,T]\times(0,\infty)).
\end{equation*}

The next result characterizes the value function and the \(q\)-function
of a fixed policy pair.
\begin{theorem}\label{thm:martingale-characterization}
Fix an admissible scale-invariant policy pair \((\pi,\eta)\) with full
support and joint density \(\nu_t(u,c)=\pi_t(u)\eta_t(c)\).  Let
\(V\in\mathcal V\) and
\(q:[0,T)\times(0,\infty)\times\mR^m\times(0,\infty)\to\mR\) be candidate
functions.  Suppose that \(J_U^{\pi,\eta}\in\mathcal V\) and that
\(q_U^{\pi,\eta}\) in \eqref{eq:little-q-definition} and \(q\) are continuous.
Assume that the localized short-time Dynkin formula associated with
\eqref{eq:little-q-definition} is valid with an integrable limit, that all
displayed terms are integrable, and that the It\^o stochastic integrals
associated with \(V\) and \(J_U^{\pi,\eta}\) are uniformly integrable
martingales for every initial pair and grid.  Assume further that
\(J_U^{\pi,\eta}\) is the unique solution in \(\mathcal V\) of the linear
Feynman--Kac terminal equation associated with \(\mathcal L^{\pi,\eta}\) and
the fixed-policy payoff.  Then
\((V,q)=(J_U^{\pi,\eta},q_U^{\pi,\eta})\) if and only if all of the following
conditions hold:
\begin{enumerate}
\item[(i)] The terminal condition is
\begin{equation}
  V(T,x)=U(x),\qquad x>0.
  \label{eq:q-terminal}
\end{equation}
\item[(ii)] For every \(t<T\) and \(x>0\),
\begin{equation}
  \int_{\mR^m}\int_0^\infty
  \left[q(t,x,u,c)-\Lambda(t,x)\log\nu_t(u,c)\right]
  \nu_t(u,c)\,\md c\,\md u=0.
  \label{eq:q-normalization}
\end{equation}
\item[(iii)] For every initial pair \((t,x)\) and every finite grid
\(\mathcal G\), the process
\begin{equation}
  M_s^{\mathcal G}[V,q]
  :=V(s,X_s^{\mathcal G})
  +\int_t^s\left[
  U(X_a^{\mathcal G}C_a^{\mathcal G})
  -q(a,X_a^{\mathcal G},A_a^{\mathcal G},C_a^{\mathcal G})
  \right]\,\md a
  \label{eq:q-martingale}
\end{equation}
is a uniformly integrable \(\mathbb F^{\mathcal G}\)-martingale on
\([t,T]\).
\end{enumerate}
\end{theorem}

\begin{proof}
See Appendix~\ref{app:proofs-policy-improvement}.
\end{proof}

The grid-sampled construction above and the requirement that the martingale
condition hold for every finite grid follow the joint characterization of the value function and
the \(q\)-function in the official erratum to \citet{JiaZhou2023}.  For jointly parameterized
\((V^\theta,q^{\theta,\psi})\), each
observed interval therefore gives
\begin{align}
  \delta_i(\theta,\psi)
  :={}&V^\theta(t_{i+1},X_{t_{i+1}}^{\mathcal G})
  -V^\theta(t_i,X_{t_i}^{\mathcal G})\notag\\
  &+\int_{t_i}^{t_{i+1}}\left[
  U(X_s^{\mathcal G}C_i)
  -q^{\theta,\psi}(s,X_s^{\mathcal G},A_i,C_i)
  \right]\,\md s .
  \label{eq:q-moment-increment}
\end{align}
The estimable policy-evaluation criterion is
\begin{equation}
  \mE\left[\zeta_i\,\delta_i(\theta,\psi)\right]=0,
  \label{eq:q-moment-condition}
\end{equation}
where \(\zeta_i\) is bounded and measurable with respect to
\(\mathcal F_{t_i}^{\mathcal G}\).  Under the usual full-rank condition
on the population moment Jacobian, a sufficiently rich collection of test functions
\(\zeta_i\) locally identifies the coefficients.  Episode averages and
quadrature give the sample analogue, following the martingale-orthogonality
approach of \citet{JiaZhou2022PE}.

The analytical results in Sections~\ref{sec:crra}--\ref{sec:log} motivate the following parametrization forms:
\begin{align}
  V_\gamma^\theta(t,x)
  &=\frac{h^\theta(t)}{1-\gamma}x^{1-\gamma},
  \qquad h^\theta(T)=1,\quad h^\theta(t)>0,
  \label{eq:data-crra-critic}\\
  q_\gamma^{\theta,\psi}(t,x,u,c)
  &=x^{1-\gamma}\left[d_\gamma^\psi(t)+\beta_\gamma^\psi(t)^\top u
  -\frac12u^\top K_\gamma^\psi(t)u
  +\frac{c^{1-\gamma}}{1-\gamma}-h^\theta(t)c\right],
  \label{eq:data-crra-q}\\
  V_{\log}^\theta(t,x)
  &=g(t)\log x+k^\theta(t),
  \qquad k^\theta(T)=0,
  \label{eq:data-log-critic}\\
  q_{\log}^{\psi}(t,x,u,c)
  &=d_{\log}^\psi(t)+\beta_{\log}^\psi(t)^\top u
  -\frac12u^\top K_{\log}^\psi(t)u
  +\log c-g(t)c,
  \label{eq:data-log-q}
\end{align}
where \(K_U^\psi(t)\) is constrained to be symmetric positive definite.  Here
\(\theta\) parameterizes \(h\) or \(k\), whereas \(\psi\) parameterizes
\(d_U,\beta_U,K_U\).  The \(q\) approximator carries both superscripts
because in the CRRA case it shares \(h^\theta\) with the value function.  The
normalization \eqref{eq:q-normalization} at \(x=1\) with \(q=q_U^{\theta,\psi}\)
determines \(d_U^\psi(t)\) under the current policy pair.  The remaining
coefficients \((h,k,\beta_U,K_U)\) are fitted through
\eqref{eq:q-moment-condition}, which does not rely on the unknown market parameters.

\subsection{Constrained policy improvement}

The next result shows that the constrained soft update improves the
current policy pair.
\begin{theorem}\label{thm:policy-improvement}
Fix an admissible scale-invariant policy pair \((\pi,\eta)\) with full
support.  Let \(J_U^{\pi,\eta}\in\mathcal V\) and \(q_U^{\pi,\eta}\) be its
value function and exact \(q\) function, respectively.  Assume that the
hypotheses of Theorem~\ref{thm:martingale-characterization} hold with
\(V=J_U^{\pi,\eta}\) and \(q=q_U^{\pi,\eta}\).  Suppose that
\(q_U^{\pi,\eta}\) has the form given in
\eqref{eq:data-crra-q} or \eqref{eq:data-log-q}, with parameter superscripts
suppressed.  Assume that \(K_U(t)\) is symmetric positive definite and, in the
CRRA case, that \(h(t)>0\).

For each \(t\in[0,T)\), define the portfolio update by
\begin{equation}
  m_U^+(t)
  :=\argmin_{m\in\mR_+^m}
  \left\{\frac12m^\top K_U(t)m-\beta_U(t)^\top m\right\},
  \qquad
  \pi_U^+(\cdot\mid t)
  :=\N\left(m_U^+(t),\Lambda(t,1)K_U(t)^{-1}\right),
  \label{eq:data-portfolio-update}
\end{equation}
and define the consumption update by
\begin{align}
  \eta_\gamma^+(c\mid t)
  &:=\frac{1}{Z_\gamma(h(t),\lambda(t))}
  \exp\left\{\frac{1}{\lambda(t)}
  \left(\frac{c^{1-\gamma}}{1-\gamma}-h(t)c\right)\right\},
  \label{eq:data-crra-consumption-update}\\
  \eta_{\log}^+(\cdot\mid t)
  &:=\operatorname{Gamma}\left(
  1+\frac{1}{\lambda},\frac{g(t)}{\lambda}\right).
  \label{eq:data-log-consumption-update}
\end{align}
Let \(\eta_U^+\) denote the corresponding update for the utility under
consideration.  If \((\pi_U^+,\eta_U^+)\) is admissible and the Dynkin formula
for \(J_U^{\pi,\eta}\) is valid under the updated policy pair for every initial
time--state pair, all terms are integrable, and its stochastic integral is
a uniformly integrable martingale, then
\begin{equation}
  J_U^{\pi_U^+,\eta_U^+}(t,x)\ge J_U^{\pi,\eta}(t,x),
  \qquad (t,x)\in[0,T]\times(0,\infty).
\end{equation}
\end{theorem}

\begin{proof}
See Appendix~\ref{app:proofs-policy-improvement}.
\end{proof}

Theorem~\ref{thm:policy-improvement} shows that a single exact update
improves the current policy. For a single risky asset and logarithmic
utility of terminal wealth, \citet[Theorem~3.5 and Appendix~A.3]{ChauNguyenNguyen2026}
recover the optimal portfolio in one update from a Gaussian initial
policy. Extending this one-step property to our joint portfolio--consumption
setting under the portfolio-mean constraint, the next theorem establishes the
convergence of both the exploratory policy densities and their deployed modes.

\begin{theorem}\label{thm:exact-policy-iteration-convergence}
Start from an admissible scale-invariant policy pair $(\pi_0,\eta_0)$
whose value $J_0$ solves its fixed-policy Feynman--Kac equation
classically. At each iteration apply
\eqref{eq:data-portfolio-update}--\eqref{eq:data-log-consumption-update}
using the exact $q$ function of the current pair. Extend the updated
densities to $t=T$ by continuity. For CRRA utility, assume the
hypotheses of Theorem~\ref{thm:crra-verification} and that
$t\mapsto(1-\gamma)J_0(t,1)$ is positive and $C^1$ on $[0,T]$.

\begin{enumerate}
\item[(i)] \emph{Logarithmic utility.}
For every $n\ge1$, $(\pi_n,\eta_n)=(\pi_{\log}^*,\eta_{\log}^*)$.
Their modes are $\widehat\mu$ and $(1+T-t)^{-1}$, respectively.

\item[(ii)] \emph{CRRA policy densities.}
There are constants $C,\alpha>0$, depending only on the model and the
initial policy, such that for every $n\ge1$,
\begin{equation}
 \sup_{t\in[0,T]}\left(
   \|\pi_n(\cdot\mid t)-\pi_\gamma^*(\cdot\mid t)\|_1
  +\|\eta_n(\cdot\mid t)-\eta_\gamma^*(\cdot\mid t)\|_1
 \right)\le C\exp(-\alpha 2^{n-1}).
 \label{eq:pi-crra-density-convergence}
\end{equation}
Here $\|\cdot\|_1$ denotes the $L^1$ norm on the respective action space.

\item[(iii)] \emph{CRRA policy modes.}
From the first update onward, the portfolio mode equals its optimal value
$\widehat\mu/\gamma$. Let $m_n(t)$ and $m^*(t)$ denote the modes of
$\eta_n(\cdot\mid t)$ and $\eta_\gamma^*(\cdot\mid t)$, respectively.
With the same constants as in \eqref{eq:pi-crra-density-convergence},
\begin{equation}
 \sup_{t\in[0,T]}|m_n(t)-m^*(t)|\le C\exp(-\alpha 2^{n-1}),
 \qquad n\ge1.
 \label{eq:pi-crra-mode-convergence}
\end{equation}
\end{enumerate}
\end{theorem}

\begin{proof}
See Appendix~\ref{app:proofs-policy-improvement}.
\end{proof}

\begin{remark}\label{rem:pi-convergence-scope}
For finite-horizon entropy-regularized diffusions with control in the
drift, \citet[Theorem~2.4]{MaWangZhang2026} establish superexponential
convergence of policy iteration. Here, CRRA homogeneity reduces each
policy-evaluation step to a scalar Newton iteration, in line with the
linearization of the Hamiltonian in their equation~(2.8).
The remainder in this step controls
the relative entropies of the updated policies and yields the density
bound \eqref{eq:pi-crra-density-convergence}.
\end{remark}

\begin{remark}\label{rem:log-policy-improvement-implementation}
For logarithmic utility, the Gamma consumption policy in
\eqref{eq:data-log-consumption-update} depends only on $g(t)$ and
$\lambda$. We therefore keep this policy fixed in the experiments of
Section~\ref{sec:experiments} and use the estimated $q$ coefficients
to update the portfolio policy.
\end{remark}

Algorithm~\ref{alg:constrained-policy-iteration} fits the value and
$q$ coefficients from sampled trajectories using \eqref{eq:q-moment-condition}
and the normalization \eqref{eq:q-normalization} at \(x=1\). It then uses
these coefficients in \eqref{eq:data-portfolio-update}--\eqref{eq:data-log-consumption-update}.
The portfolio update enforces the nonnegative-mean constraint through
the quadratic program in \eqref{eq:data-portfolio-update}.

\begin{algorithm}[H]
\caption{Constrained martingale policy iteration with modal deployment}
\label{alg:constrained-policy-iteration}
\begin{algorithmic}[1]
\Require Utility \(U\in\{U_\gamma,\log\}\), temperature \(\Lambda(\cdot,1)\), grid
  \(\mathcal G\), number of episodes \(L\ge1\), initial policy pair
  \((\pi_U^{(0)},\eta_U^{(0)})\)
\For{\(n=0,\ldots,L-1\)}
  \State Set
  \(\nu_t^{(n)}(u,c)\gets
    \pi_U^{(n)}(u\mid t)\eta_U^{(n)}(c\mid t)\).
  \State Under \(\nu^{(n)}\), sample \((A_i,C_i)\) on \(\mathcal G\) and
  observe the wealth trajectory \(X^{\mathcal G}\) in
  \eqref{eq:sampled-wealth}.
  \State Fit \((\widehat\theta_n,\widehat\psi_n)\) using
  \eqref{eq:q-terminal}, \eqref{eq:q-normalization} at \(x=1\) under
  \(\nu^{(n)}\), and the sample
  analogues of \eqref{eq:q-moment-condition}.
  \State Let \(\widehat\beta_{U,n}:=\beta_U^{\widehat\psi_n}\),
  \(\widehat K_{U,n}:=K_U^{\widehat\psi_n}\), and, for CRRA utility,
  \(\widehat h_n:=h^{\widehat\theta_n}\).
  \ForAll{\(t\in[0,T)\)}
    \State \(m_U^{(n+1)}(t)\gets
      \displaystyle\argmin_{m\in\mR_+^m}
      \left\{\frac12m^\top\widehat K_{U,n}(t)m
      -\widehat\beta_{U,n}(t)^\top m\right\}\).
    \State \(\pi_U^{(n+1)}(\cdot\mid t)\gets
      \N\left(m_U^{(n+1)}(t),
      \Lambda(t,1)\widehat K_{U,n}(t)^{-1}\right)\).
    \If{\(U=U_\gamma\)}
      \State Set \(\eta_U^{(n+1)}(\cdot\mid t)\) to the Gibbs density in
      \eqref{eq:data-crra-consumption-update} with \(h=\widehat h_n\).
    \Else
      \State \(\eta_U^{(n+1)}(\cdot\mid t)\gets
      \operatorname{Gamma}\left(1+1/\lambda,g(t)/\lambda\right)\).
    \EndIf
  \EndFor
\EndFor
\State Deploy
\(u_U^{\mathrm{dep}}(t)\gets
  \argmax_{u\in\mR^m}\pi_U^{(L)}(u\mid t)=m_U^{(L)}(t)\in\mR_+^m\).
\If{\(U=U_\gamma\)}
  \State Deploy
  \(c_U^{\mathrm{dep}}(t)\gets
  \argmax_{c>0}\eta_U^{(L)}(c\mid t)
  =\widehat h_{L-1}(t)^{-1/\gamma}\).
\Else
  \State Deploy
  \(c_U^{\mathrm{dep}}(t)\gets
  \argmax_{c>0}\eta_U^{(L)}(c\mid t)=1/g(t)\).
\EndIf
\end{algorithmic}
\end{algorithm}

The algorithm deploys the modes of the fitted densities. For the optimal
densities, Proposition~\ref{prop:modal-recovery} identifies the portfolio
modes and the logarithmic consumption mode with the corresponding
classical controls.

\section{Numerical Experiments}\label{sec:experiments}

We examine the logarithmic specification of
Algorithm~\ref{alg:constrained-policy-iteration} in two constant-coefficient,
two-asset markets.  The first has an interior oracle portfolio and is used to
assess coefficient recovery.  In the second, both assets have positive excess
returns, yet the unconstrained optimum shorts one of them; this makes the
no-shorting boundary economically nontrivial.  The comparison concerns the
learned little-$q$ coefficients, the modal portfolio produced by policy
improvement, and its out-of-sample objective.  As in the formulation of
Section~\ref{sec:model}, the nonnegativity constraint applies to the Gaussian mean (and hence
its mode), yet individual exploratory actions may still take negative values.

\paragraph{Market models.}
We consider two risky assets over $T=1$, with initial wealth $x_0=1$, risk-free
rate $r=0.03$, and risky-asset return vector
\begin{equation}
  b=\begin{pmatrix}0.10\\0.18\end{pmatrix}.
  \label{eq:experiment-market}
\end{equation}
Thus the excess-return vector is
$e=b-r\mathbf{1}_2=(0.07,0.15)^\top$.  To illustrate the effect of no-shorting constraints in the multi-asset setting, we consider the following two different covariance matrices:
\begin{equation}
  \Sigma^{\mathrm I}
  =\begin{pmatrix}0.040&0.006\\0.006&0.090\end{pmatrix},
  \qquad
  \Sigma^{\mathrm B}
  =\begin{pmatrix}0.040&0.045\\0.045&0.090\end{pmatrix},
  \label{eq:experiment-covariance-designs}
\end{equation}
where the superscripts $\mathrm{I}$ and $\mathrm{B}$ refer to the ``interior'' and ``boundary'' markets, respectively.

\paragraph{Interior design: inactive constraint.}
For $\Sigma^{\mathrm I}$, solving the quadratic program in
\eqref{eq:data-portfolio-update} with the true coefficients gives
\begin{equation}
  \widehat\mu^{\mathrm I}
  =\argmin_{m\in\mR_+^2}
  \left\{\frac12m^\top\Sigma^{\mathrm I}m-e^\top m\right\}
  =\begin{pmatrix}1.515152\\1.565657\end{pmatrix}.
  \label{eq:experiment-interior-oracle-mode}
\end{equation}
Both components are strictly positive, so the constraint is inactive at this
oracle even though every policy-improvement step still solves the nonnegative
quadratic program.

\paragraph{Boundary design: active no-shorting constraint.}
The second design changes only the off-diagonal covariance.  The two assets
retain positive individual Sharpe ratios, $0.07/0.20=0.35$ and
$0.15/0.30=0.50$, while their correlation becomes $0.75$.  Nevertheless, the
unconstrained portfolio and the constrained oracle are, respectively,
\begin{equation}
  (\Sigma^{\mathrm B})^{-1}e
  =\begin{pmatrix}-0.285714\\1.809524\end{pmatrix},
  \qquad
  \widehat\mu^{\mathrm B}
  =\argmin_{m\in\mR_+^2}
  \left\{\frac12m^\top\Sigma^{\mathrm B}m-e^\top m\right\}
  =\begin{pmatrix}0\\1.666667\end{pmatrix}.
  \label{eq:experiment-boundary-oracle-mode}
\end{equation}
Thus the unconstrained sign pattern is $(-,+)$, whereas the admissible oracle
has sign pattern $(0,+)$.  The second component also changes from $1.809524$ to
$1.666667$, because the constrained optimum is obtained by resolving the
quadratic program rather than by truncating its negative component.
Indeed, at $\widehat\mu^{\mathrm B}$ the first-coordinate KKT multiplier is
$0.045(1.666667)-0.07=0.005>0$, whereas the second-coordinate stationarity
condition holds with equality.

We set $\lambda=0.1$ and use $N=252$ equal time intervals.  Under the oracle
logarithmic policy in the interior design, the exploratory portfolio covariance
at time zero is
\begin{equation}
  \frac{\lambda}{g(0)}(\Sigma^{\mathrm I})^{-1}
  =\begin{pmatrix}
     1.262626&-0.084175\\
    -0.084175& 0.561167
   \end{pmatrix}.
  \label{eq:experiment-interior-oracle-covariance}
\end{equation}
As explained in Remark~\ref{rem:log-policy-improvement-implementation}, we keep
the logarithmic consumption actor fixed at the explicit update
\eqref{eq:data-log-consumption-update} across policy iterations.  With
\(\lambda=0.1\), this actor is \(\operatorname{Gamma}(11,g(t)/0.1)\), whose
mode is $1/g(t)$; the deployed relative consumption rate therefore starts at
$1/2$ and approaches one at maturity.

\paragraph{Numerical implementation of policy evaluation.}
For the logarithmic specification, we use
\begin{equation}
  q_{\log}(t,u,c)
  =d(t)+g(t)(b^q)^\top u
  -\frac{g(t)}2u^\top K^q u+\log c-g(t)c.
  \label{eq:experiment-q-class}
\end{equation}
In this specification the grid transition of log wealth gives
\begin{equation}
  \frac{\log X_{t_{i+1}}-\log X_{t_i}}{\Delta t}+C_i
  =r+(b^q)^\top A_i-\frac12A_i^\top K^q A_i+\varepsilon_i,
  \qquad \mE[\varepsilon_i\mid\mathcal F^{\mathcal G}_{t_i},A_i,C_i]=0.
  \label{eq:experiment-transition-regression}
\end{equation}
We therefore regress the left-hand side on
\begin{equation*}
  1,\quad A_{i,1},\quad A_{i,2},\quad
  -\frac12A_{i,1}^2,\quad-A_{i,1}A_{i,2},\quad
  -\frac12A_{i,2}^2.
\end{equation*}
The intercept absorbs the risk-free rate and is not used in policy
improvement.  We regard \eqref{eq:experiment-transition-regression} as the
constant-coefficient logarithmic specialization of the martingale evaluation
criterion in Section \ref{sec:policy-improvement}.  It permits $(b^q,K^q)$ to
be estimated before the value coefficients.  Although the behavior policy
changes from one pass to the next, the population regression coefficients remain
$b^q=e$ and $K^q=\Sigma$; their sufficient statistics can therefore be
accumulated across passes.  The time-dependent coefficients in
\eqref{eq:data-log-q} are recovered as $g(t)b^q$ and $g(t)K^q$.  A small ridge
is used when solving the sample normal equations, and the symmetric estimate
of $K^q$ is projected onto the positive-definite cone before policy
improvement.

Given $(b^q,K^q)$ and the current actor, the normalization
\eqref{eq:q-normalization} at $x=1$ determines $d(t)$ analytically, with both
Gaussian and Gamma entropy included.  For the value function we use
\begin{equation}
  V^\theta_{\log}(t,x)=g(t)\log x+k^\theta(t),
  \qquad
  k^\theta(t)=\theta_1\tau+\theta_2\tau^2
  -\left[1+\lambda\left(1+\frac m2\right)\right]
  \big[(1+\tau)\log(1+\tau)-(1+\tau)+1\big],
  \label{eq:experiment-value-class}
\end{equation}
where $\tau=T-t$.  This class satisfies the terminal condition and contains the
fixed-policy value function for the model under consideration.  Its two free
coefficients are fitted using a left-endpoint discretization of
\eqref{eq:q-martingale}.  Unlike $(b^q,K^q)$, they depend on the current actor,
so only the current batch is used.

\paragraph{Sampling schedule.}
The initial portfolio actor has mean zero and precision $0.1I_2$, and hence
full support with broad dispersion.  All $5{,}000$ trajectories in the first
policy-evaluation batch enter the cumulative sufficient statistics for
$(b^q,K^q)$ and count toward the reported training budget.  The first point in each convergence
figure is the policy obtained from this batch; the unestimated initial actor at
episode zero is not plotted.  Subsequent batches contain $250$ trajectories and
replace the actor by
\begin{equation*}
  m^+=\argmin_{m\in\mR_+^2}
  \left\{\frac12m^\top K^qm-(b^q)^\top m\right\},
  \qquad
  \pi^+(\cdot\mid t)=
  \N\left(m^+,\frac{\lambda}{g(t)}(K^q)^{-1}\right).
\end{equation*}
The fixed training budget is $150{,}000$ trajectories, corresponding to $581$
policy passes.  Conditional on the actor for a given pass, trajectories within
the batch are independent.  Portfolio and consumption actions are redrawn at
each grid point, independently of future Brownian increments, and held over the
next interval.  Both designs use seed $20260520$ and otherwise identical
settings.

\begin{figure}[htbp]
  \centering
  \includegraphics[width=0.98\linewidth]
    {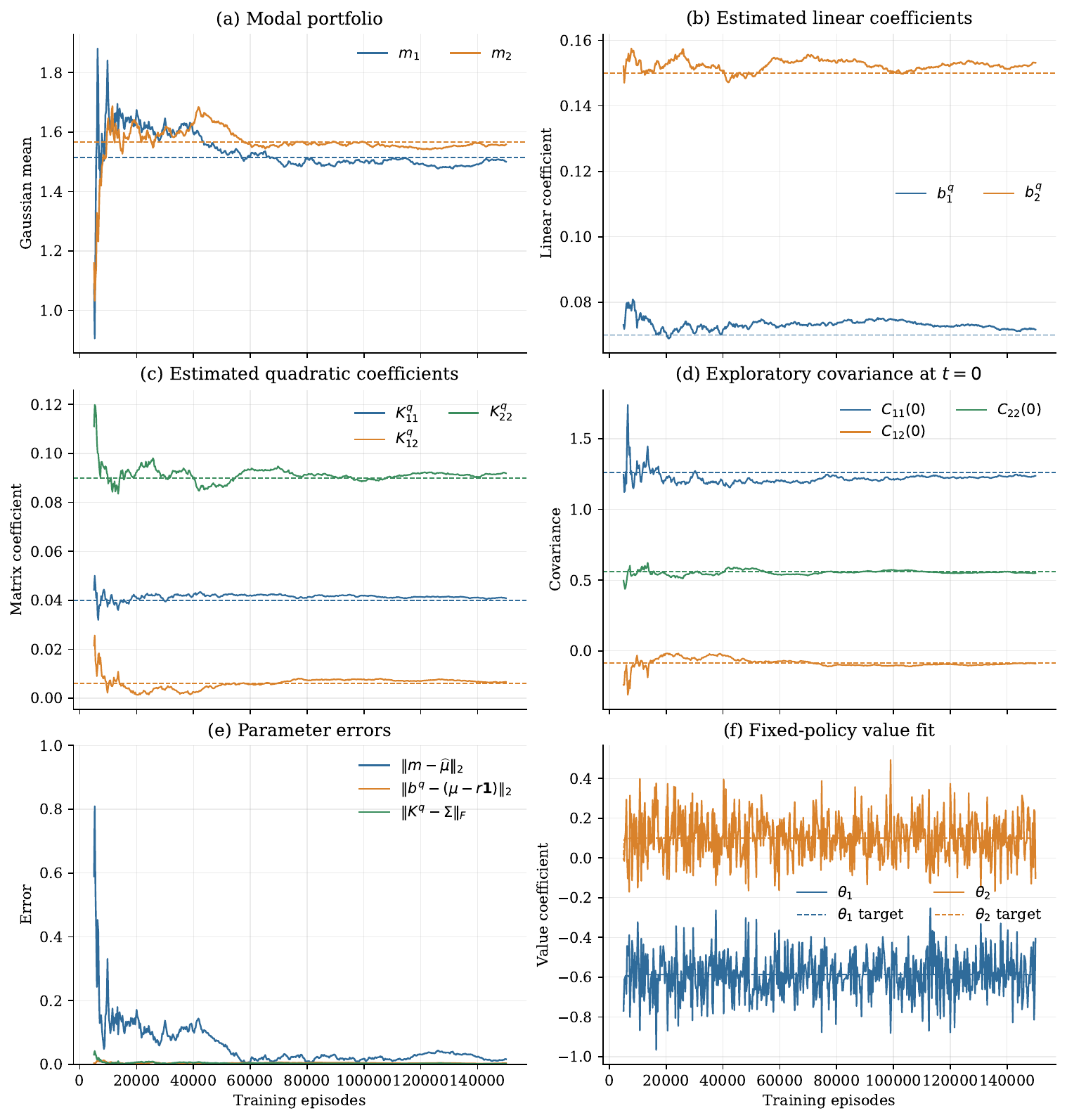}
  \caption{Constrained martingale policy iteration in the interior design.
  Panels show the Gaussian mode, estimated linear coefficients, estimated
  quadratic coefficients, exploratory covariance at $t=0$, parameter errors,
  and current-policy value coefficients.  Solid lines are estimates and
  dashed lines are full-information targets.}
  \label{fig:experiment-interior-convergence}
\end{figure}

\begin{figure}[htbp]
  \centering
  \includegraphics[width=0.98\linewidth]
    {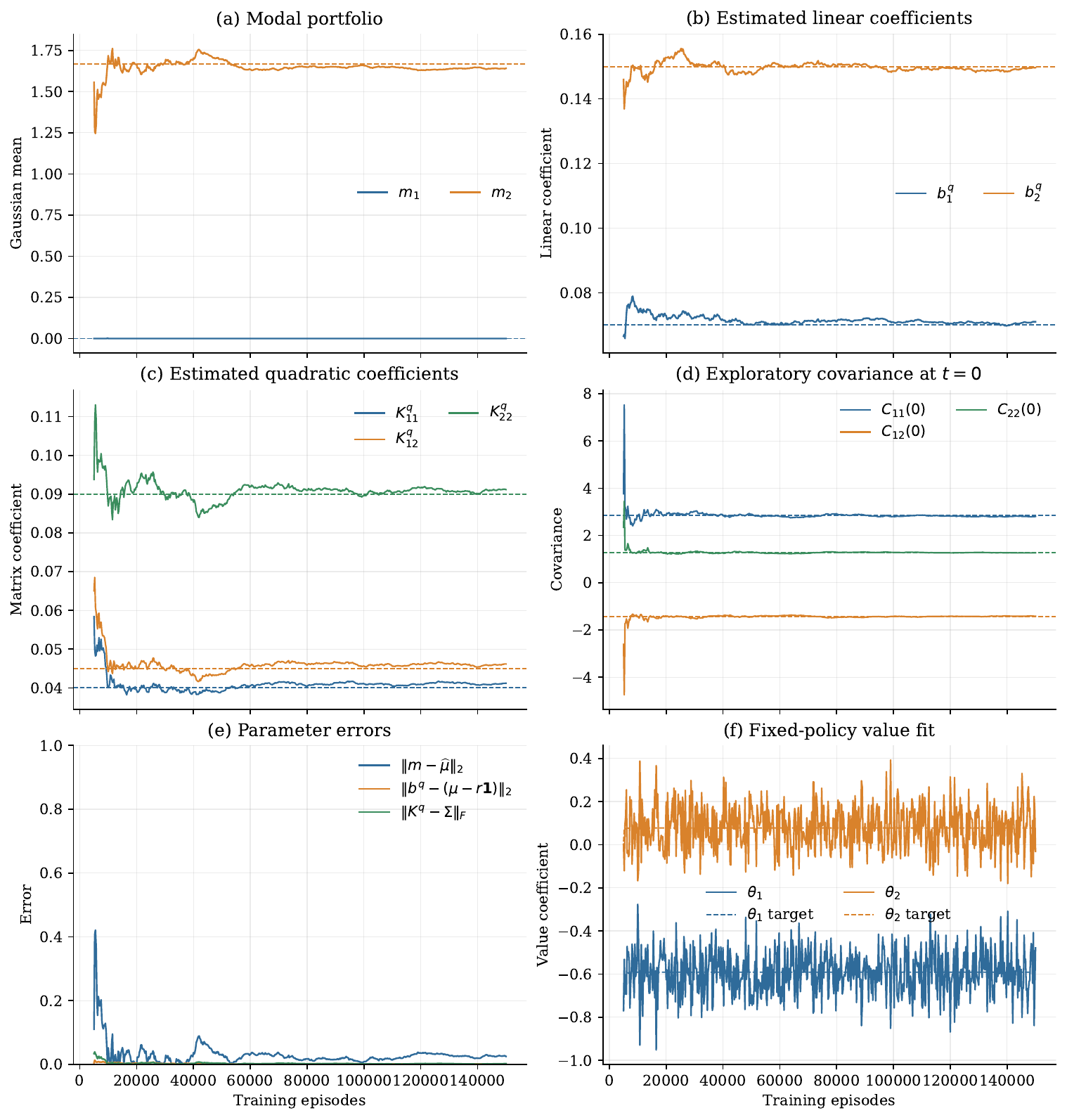}
  \caption{Constrained martingale policy iteration in the active-constraint
  design.  Its unconstrained oracle has sign pattern $(-,+)$, whereas the
  no-shorting oracle and learned modal portfolio have sign pattern $(0,+)$.
  Panels (b) and (c) show the estimated linear coefficients and estimated
  quadratic coefficients, respectively.  The panels have the same definitions as in
  Figure~\ref{fig:experiment-interior-convergence}.  The first component of the
  modal portfolio is zero at the constrained oracle and, after the active set
  is identified, in the learned policy.  Solid lines are estimates and dashed
  lines are full-information targets.}
  \label{fig:experiment-boundary-convergence}
\end{figure}

\paragraph{Parameter convergence.}
Figures~\ref{fig:experiment-interior-convergence} and
\ref{fig:experiment-boundary-convergence} record the estimates after every
policy pass.  In the interior design, the Gaussian mode settles near
\eqref{eq:experiment-interior-oracle-mode} as the pooled transition sample
grows.  The estimates of $(b^q,K^q)$ approach $(e,\Sigma^{\mathrm I})$, and the
implied covariance approaches
\eqref{eq:experiment-interior-oracle-covariance}.  Residual nonmonotonicity is
expected because each actor replacement uses newly generated trajectories.

The boundary design gives a sharper test of policy improvement.  Across $581$
passes, $m_1=0$ except for one early estimate of $0.0030$, and the last $100$
updates all select the oracle active set; $m_2$ remains near its positive
target.  At the final estimate, the first-coordinate KKT slack is
$K^q_{12}m_2-b^q_1=0.004739>0$.

The final estimates for both designs are reported in
Table~\ref{tab:experiment-estimates}.  In the interior case, the Euclidean
errors of the mode and $b^q$ are $0.016515$ and $0.003574$, and the Frobenius
error of $K^q$ is $0.002262$.  In the boundary case, the corresponding errors
are $0.025312$, $0.001084$, and $0.002345$.  In both cases the regression
recovers the off-diagonal coefficient.

\begin{table}[t]
  \centering
  \caption{Final estimates after $150{,}000$ training trajectories in the
  interior and boundary designs.}
  \label{tab:experiment-estimates}
  \begin{tabular}{lrrrr}
    \hline
    & \multicolumn{2}{c}{Interior design}
    & \multicolumn{2}{c}{Boundary design}\\
    \cline{2-3}\cline{4-5}
    Quantity & Estimate & Target & Estimate & Target\\
    \hline
    $m_1$       & 1.500313 & 1.515152 & 0.000000 & 0.000000\\
    $m_2$       & 1.558408 & 1.565657 & 1.641355 & 1.666667\\
    $b^q_1$     & 0.071663 & 0.070000 & 0.071028 & 0.070000\\
    $b^q_2$     & 0.153163 & 0.150000 & 0.149656 & 0.150000\\
    $K^q_{11}$  & 0.040858 & 0.040000 & 0.041189 & 0.040000\\
    $K^q_{12}$  & 0.006650 & 0.006000 & 0.046161 & 0.045000\\
    $K^q_{22}$  & 0.091880 & 0.090000 & 0.091178 & 0.090000\\
    \hline
  \end{tabular}
\end{table}

Panel~(f) is more variable because the value coefficients and their diagnostic
targets change with the actor and are fitted from the current batch.  This
variability does not enter the portfolio update, which depends on the pooled
estimates of $(b^q,K^q)$; consumption is updated analytically.

\paragraph{Modal deployment.}
Deployment uses the modes of the fitted Gaussian and Gamma densities.  We
evaluate the resulting deterministic controls on $5{,}000$ fresh paths and use
common Brownian increments for the learned and oracle portfolios.  Entropy is
excluded from this comparison: the criterion is the classical logarithmic
consumption--terminal-wealth objective in Proposition~\ref{prop:modal-recovery}.

In the interior design, the learned mode $(1.500313,1.558408)^\top$ and the
oracle yield mean objectives $-1.089976$ and $-1.090042$.  Their paired
learned-minus-oracle difference is $6.57\times10^{-5}$, with standard error
$8.32\times10^{-5}$ and analytical expectation $-1.11\times10^{-5}$.  The
small sign reversal is therefore compatible with Monte Carlo variation.  For
the boundary design, the corresponding means are $-1.160688$ and $-1.160766$;
the paired difference is $7.89\times10^{-5}$, its standard error is
$1.64\times10^{-4}$, and its analytical expectation is
$-4.32\times10^{-5}$.

\begin{figure}[htbp]
  \centering
  \includegraphics[width=0.98\linewidth]
    {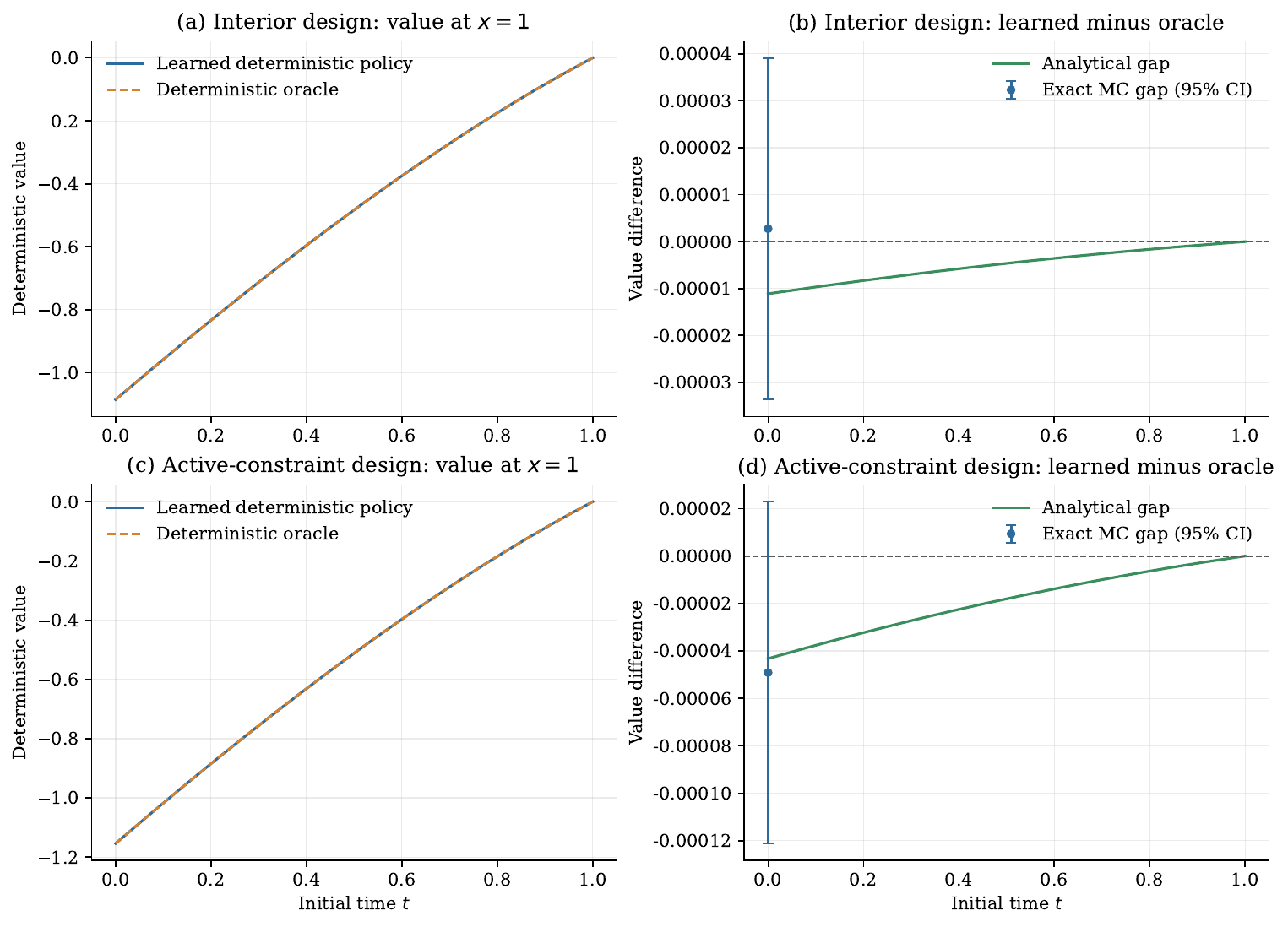}
  \caption{Value functions of the learned deterministic modal policies and the
  deterministic full-information policies at $x=1$.  The left panels show the
  value levels; the right panels show the learned-minus-oracle differences.
  Points and error bars give exact Brownian Monte Carlo estimates at $t=0$ and
  their $95\%$ confidence intervals.  No policy randomization or entropy term
  is used in this comparison.}
  \label{fig:experiment-deterministic-value}
\end{figure}

Figure~\ref{fig:experiment-deterministic-value} compares the corresponding
deterministic policy values.  At $x=1$ the learned and oracle curves overlap
visually; their magnified gaps at $t=0$ are $-1.11\times10^{-5}$ and
$-4.32\times10^{-5}$ and vanish at maturity.  Hence the parameter errors incur
only a small loss under the classical deployment objective.

\bigskip\noindent\textbf{Acknowledgements.} Xiang Yu is supported by the Hong Kong RGC General Research Fund (GRF) under grants no. 15211524 and no. 15214125. Fengyi Yuan is supported by the startup fund of the Chinese University of Hong Kong (Shenzhen) under grant no. UDF01004253.

\bibliographystyle{abbrvnat}
\bibliography{ref.bib}

\begin{thebibliography}{32}
\providecommand{\natexlab}[1]{#1}
\providecommand{\url}[1]{\texttt{#1}}
\expandafter\ifx\csname urlstyle\endcsname\relax
  \providecommand{\doi}[1]{doi: #1}\else
  \providecommand{\doi}{doi: \begingroup \urlstyle{rm}\Url}\fi

\bibitem[Allen et~al.(1993)Allen, Morris, and Postlewaite]{Allen93}
F.~Allen, S.~Morris, and A.~Postlewaite.
\newblock Finite bubbles with short sale constraints and asymmetric
  information.
\newblock \emph{Journal of Economic Theory}, 61\penalty0 (2):\penalty0
  206--229, 1993.

\bibitem[Beber and Pagano(2013)]{Beber}
A.~Beber and M.~Pagano.
\newblock Short-selling bans around the world: Evidence from the 2007--09
  crisis.
\newblock \emph{The Journal of Finance}, 68\penalty0 (1):\penalty0 343--381,
  2013.

\bibitem[Bo et~al.(2025)Bo, Huang, and Yu]{BHY25}
L.~Bo, Y.~Huang, and X.~Yu.
\newblock On optimal tracking portfolio in incomplete markets: The
  reinforcement learning approach.
\newblock \emph{SIAM Journal on Control and Optimization}, 63\penalty0
  (1):\penalty0 321--348, 2025.

\bibitem[Bo et~al.(2026)Bo, Huang, Yu, and Zhang]{BHYZ26}
L.~Bo, Y.~Huang, X.~Yu, and T.~Zhang.
\newblock Continuous-time q-learning in jump-diffusion models under {Tsallis}
  entropy.
\newblock \emph{Mathematics and Financial Economics}, 20\penalty0 (3):\penalty0
  501--546, 2026.

\bibitem[Chau et~al.(2026)Chau, Nguyen, and Nguyen]{ChauNguyenNguyen2026}
H.~Chau, D.~Nguyen, and T.~Nguyen.
\newblock Continuous-time optimal investment with portfolio constraints: A
  reinforcement learning approach.
\newblock \emph{European Journal of Operational Research}, 328\penalty0
  (3):\penalty0 1068--1092, 2026.

\bibitem[Cvitani\'{c} and Karatzas(1992)]{Cvit}
J.~Cvitani\'{c} and I.~Karatzas.
\newblock Convex duality in constrained portfolio optimization.
\newblock \emph{The Annals of Applied Probability}, 2\penalty0 (4):\penalty0
  767--818, 1992.

\bibitem[Dai et~al.(2023)Dai, Dong, and Jia]{Dong2023}
M.~Dai, Y.~Dong, and Y.~Jia.
\newblock Learning equilibrium mean-variance strategy.
\newblock \emph{Mathematical Finance}, 33\penalty0 (4):\penalty0 1166--1212,
  2023.

\bibitem[Dai et~al.(2025)Dai, Dong, Jia, and Zhou]{RL-Merton}
M.~Dai, Y.~Dong, Y.~Jia, and X.~Y. Zhou.
\newblock Data-driven {Merton}'s strategies via policy randomization.
\newblock \emph{Preprint, arXiv:2312.11797}, 2025.

\bibitem[Frikha et~al.(2025)Frikha, Germain, Lauri\`{e}re, Pham, and
  Song]{FGLS}
N.~Frikha, M.~Germain, M.~Lauri\`{e}re, H.~Pham, and X.~Song.
\newblock Actor-critic learning for mean-field control in continuous time.
\newblock \emph{Journal of Machine Learning Research}, 26\penalty0
  (127):\penalty0 1--42, 2025.

\bibitem[Gao et~al.(2026)Gao, Li, and Zhou]{GLZ26}
X.~Gao, L.~Li, and X.~Y. Zhou.
\newblock Reinforcement learning for jump-diffusions, with financial
  applications.
\newblock \emph{Mathematical Finance}, pages 1--32, 2026.
\newblock Early View.

\bibitem[Giegrich et~al.(2024)Giegrich, Reisinger, and Zhang]{GRZ}
M.~Giegrich, C.~Reisinger, and Y.~Zhang.
\newblock Convergence of policy gradient methods for finite-horizon exploratory
  linear-quadratic control problems.
\newblock \emph{SIAM Journal on Control and Optimization}, 62\penalty0
  (2):\penalty0 1060--1092, 2024.

\bibitem[G\'{o}mez and Sharma(2006)]{Gomez}
J.-P. G\'{o}mez and T.~Sharma.
\newblock Portfolio delegation under short-selling constraints.
\newblock \emph{Economic Theory}, 28\penalty0 (1):\penalty0 173--196, 2006.

\bibitem[Grullon et~al.(2015)Grullon, Michenaud, and Weston]{Grullon}
G.~Grullon, S.~Michenaud, and J.~P. Weston.
\newblock The real effects of short-selling constraints.
\newblock \emph{The Review of Financial Studies}, 28\penalty0 (6):\penalty0
  1737--1767, 2015.

\bibitem[Guo et~al.(2022)Guo, Xu, and Zariphopoulou]{GXZ22}
X.~Guo, R.~Xu, and T.~Zariphopoulou.
\newblock Entropy regularization for mean field games with learning.
\newblock \emph{Mathematics of Operations Research}, 47\penalty0 (4):\penalty0
  3239--3260, 2022.

\bibitem[Jia and Zhou(2022{\natexlab{a}})]{JiaZhou2022PE}
Y.~Jia and X.~Y. Zhou.
\newblock Policy evaluation and temporal-difference learning in continuous time
  and space: A martingale approach.
\newblock \emph{Journal of Machine Learning Research}, 23\penalty0
  (154):\penalty0 1--55, 2022{\natexlab{a}}.

\bibitem[Jia and Zhou(2022{\natexlab{b}})]{jia2022policy}
Y.~Jia and X.~Y. Zhou.
\newblock Policy gradient and actor-critic learning in continuous time and
  space: Theory and algorithms.
\newblock \emph{Journal of Machine Learning Research}, 23\penalty0
  (275):\penalty0 1--50, 2022{\natexlab{b}}.

\bibitem[Jia and Zhou(2023)]{JiaZhou2023}
Y.~Jia and X.~Y. Zhou.
\newblock {q}-learning in continuous time.
\newblock \emph{Journal of Machine Learning Research}, 24\penalty0
  (161):\penalty0 1--61, 2023.

\bibitem[Jia et~al.(2026)Jia, Ouyang, and Zhang]{JiaOuyangZhang2025}
Y.~Jia, D.~Ouyang, and Y.~Zhang.
\newblock Accuracy of discretely sampled stochastic policies in continuous-time
  reinforcement learning.
\newblock \emph{SIAM Journal on Control and Optimization}, 64\penalty0
  (3):\penalty0 1889--1929, 2026.

\bibitem[Li et~al.(2026)Li, Wang, and Zhou]{LiWangZhou2026}
X.~Li, Y.~Wang, and X.~Y. Zhou.
\newblock Dynamic mean--variance portfolio selection with no-shorting
  constraints and unknown investment opportunity sets.
\newblock \emph{Preprint, arXiv:2607.16625}, 2026.

\bibitem[Ma et~al.(2026)Ma, Wang, and Zhang]{MaWangZhang2026}
J.~Ma, G.~Wang, and J.~Zhang.
\newblock Convergence analysis for entropy-regularized control problems: A
  probabilistic approach.
\newblock \emph{SIAM Journal on Control and Optimization}, 64\penalty0
  (2):\penalty0 816--842, 2026.
\newblock \doi{10.1137/24M1680039}.

\bibitem[Pham(2009)]{Pham2009}
H.~Pham.
\newblock \emph{Continuous-Time Stochastic Control and Optimization with
  Financial Applications}, volume~61 of \emph{Stochastic Modelling and Applied
  Probability}.
\newblock Springer, Berlin, 2009.

\bibitem[Shirakawa(1994)]{Shirakawa}
H.~Shirakawa.
\newblock Optimal consumption and portfolio selection with incomplete markets
  and upper and lower bound constraints.
\newblock \emph{Mathematical Finance}, 4\penalty0 (1):\penalty0 1--24, 1994.

\bibitem[Szpruch et~al.(2024)Szpruch, Treetanthiploet, and
  Zhang]{szpruch2024optimal}
L.~Szpruch, T.~Treetanthiploet, and Y.~Zhang.
\newblock Optimal scheduling of entropy regularizer for continuous-time
  linear-quadratic reinforcement learning.
\newblock \emph{SIAM Journal on Control and Optimization}, 62\penalty0
  (1):\penalty0 135--166, 2024.

\bibitem[Tang et~al.(2022)Tang, Zhang, and Zhou]{TangZhangZhou2022}
W.~Tang, Y.~P. Zhang, and X.~Y. Zhou.
\newblock Exploratory {HJB} equations and their convergence.
\newblock \emph{SIAM Journal on Control and Optimization}, 60\penalty0
  (6):\penalty0 3191--3216, 2022.

\bibitem[Teschl(2012)]{Teschl2012}
G.~Teschl.
\newblock \emph{Ordinary Differential Equations and Dynamical Systems}, volume
  140 of \emph{Graduate Studies in Mathematics}.
\newblock American Mathematical Society, Providence, RI, 2012.

\bibitem[Wang et~al.(2026)Wang, Gao, and Li]{WGL26}
B.~Wang, X.~Gao, and L.~Li.
\newblock Reinforcement learning for continuous-time optimal execution:
  actor-critic algorithm and error analysis.
\newblock \emph{Finance and Stochastics}, 30\penalty0 (2):\penalty0 597--655,
  2026.

\bibitem[Wang et~al.(2020)Wang, Zariphopoulou, and
  Zhou]{WangZariphopoulouZhou2020}
H.~Wang, T.~Zariphopoulou, and X.~Y. Zhou.
\newblock Reinforcement learning in continuous time and space: A stochastic
  control approach.
\newblock \emph{Journal of Machine Learning Research}, 21\penalty0
  (198):\penalty0 1--34, 2020.

\bibitem[Wang et~al.(2024)Wang, Yan, and Yu]{WYY}
W.~Wang, K.~Yan, and X.~Yu.
\newblock Optimal portfolio under ratio-type periodic evaluation in stochastic
  factor models under convex trading constraints.
\newblock \emph{Preprint, arXiv:2411.13579}, 2024.

\bibitem[Wei and Yu(2025)]{WY25}
X.~Wei and X.~Yu.
\newblock Continuous time {q}-learning for mean-field control problems.
\newblock \emph{Applied Mathematics \& Optimization}, 91\penalty0 (1):\penalty0
  10, 2025.

\bibitem[Wei et~al.(2024)Wei, Yu, and Yuan]{WYY24}
X.~Wei, X.~Yu, and F.~Yuan.
\newblock Unified continuous-time q-learning for mean-field game and mean-field
  control problems.
\newblock \emph{Preprint, arXiv:2407.04521}, 2024.

\bibitem[Wu and Li(2024)]{wu2024reinforcement}
B.~Wu and L.~Li.
\newblock Reinforcement learning for continuous-time mean-variance portfolio
  selection in a regime-switching market.
\newblock \emph{Journal of Economic Dynamics and Control}, 158:\penalty0
  104787, 2024.

\bibitem[Xu and Shreve(1992)]{XS1992}
G.-L. Xu and S.~E. Shreve.
\newblock {A Duality Method for Optimal Consumption and Investment Under
  Short-Selling Prohibition. II. Constant Market Coefficients}.
\newblock \emph{The Annals of Applied Probability}, 2\penalty0 (2):\penalty0
  314--328, 1992.

\end{thebibliography}

\appendix

\section{Constrained entropy-regularized quadratic maximization}

This appendix establishes an auxiliary lemma for constrained entropy-regularized quadratic maximization problems.  For a symmetric positive-definite matrix $A\in\mR^{m\times m}$,
$p\in\mR^m$, and $\alpha>0$, consider
\begin{equation}
  \sup_{\pi\in\cP_\pi(\mR^m)}
  \left\{
  \int_{\mR^m}\left(p^\top u-\frac12u^\top Au\right)
  \pi(u)\,\md u+\alpha\cH(\pi)
  \right\}.
  \label{eq:quadratic-entropy-problem}
\end{equation}
The positive definiteness of $A$ makes the finite-second-moment condition in
Definition~\ref{def:density-classes} equivalent to
$\int_{\mR^m}u^\top Au\pi(u)\,\md u<\infty$.
Write $G(\pi)$ for the objective functional in
\eqref{eq:quadratic-entropy-problem}.

\begin{lemma}\label{lem:quadratic-entropy}
\emph{(i)} Let $f$ be measurable and suppose
$0<Z_f:=\int\exp(f(z)/\alpha)\,\md z<\infty$.  If
\begin{equation}
  q^*(z):=Z_f^{-1}\exp(f(z)/\alpha),
\end{equation}
then every density $q$ satisfying
$\int |f(z)|q(z)\,\md z<\infty$ and
$\int q(z)|\log q(z)|\,\md z<\infty$ satisfies
\begin{equation}
  \alpha\log Z_f-\int f(z)q(z)\,\md z-\alpha\cH(q)
  =\alpha\operatorname{KL}(q\Vert q^*).
\end{equation}

\emph{(ii)} Let
\begin{equation}
  a^*:=\argmin_{a\in\mR_+^m}
  \left\{\frac12a^\top Aa-p^\top a\right\}.
  \label{eq:quadratic-primal}
\end{equation}
Set $\ell^*:=Aa^*-p$.  The optimality conditions for
\eqref{eq:quadratic-primal} are
\begin{equation}
  a^*\ge0,
  \qquad
  \ell^*\ge0,
  \qquad
  (\ell^*)^\top a^*=0.
  \label{eq:quadratic-kkt}
\end{equation}
Define
  $\pi^*:=\N\left(a^*,\alpha A^{-1}\right)$.
Then every $\pi\in\cP_\pi(\mR^m)$ satisfies the exact gap identity
\begin{equation}
  G(\pi^*)-G(\pi)
  =\alpha\operatorname{KL}(\pi\Vert\pi^*)
  +(\ell^*)^\top\int_{\mR^m}u\pi(u)\,\md u.
\end{equation}
Consequently, $\pi^*$ is the unique maximizer of
\eqref{eq:quadratic-entropy-problem}, whose optimal value is
\begin{equation}
  \frac12(a^*)^\top Aa^*
  +\frac{\alpha}{2}
  \log\left(\frac{(2\pi\alpha)^m}{\det(A)}\right).
\end{equation}

\end{lemma}

\begin{proof}
The identity $\log q^*=f/\alpha-\log Z_f$ gives
\begin{align}
  \alpha\operatorname{KL}(q\Vert q^*)
  &=\alpha\int q(z)\log q(z)\,\md z
  -\int f(z)q(z)\,\md z+\alpha\log Z_f\\
  &=\alpha\log Z_f-\int f(z)q(z)\,\md z-\alpha\cH(q),
\end{align}
which proves part~(i).

Strict convexity and coercivity give the unique minimizer $a^*$ in
\eqref{eq:quadratic-primal}, whose KKT conditions are
\eqref{eq:quadratic-kkt}.  Define the augmented reward
\begin{equation}
  f_*(u):=(p+\ell^*)^\top u-\frac12u^\top Au
  =\frac12(a^*)^\top Aa^*
  -\frac12(u-a^*)^\top A(u-a^*).
\end{equation}
Its partition function and Gibbs density are
\begin{align}
  &Z_*
  =\exp\left(\frac{(a^*)^\top Aa^*}{2\alpha}\right)
  \frac{(2\pi\alpha)^{m/2}}{\det(A)^{1/2}},\\
  &Z_*^{-1}\exp(f_*/\alpha)
  =\N(a^*,\alpha A^{-1})=\pi^*.
\end{align}
The Gaussian form of $\pi^*$ and the condition $a^*\ge0$ show that
$\pi^*\in\cP_\pi(\mR^m)$.  The KKT conditions also give
$(\ell^*)^\top a^*=0$, and hence
\begin{equation}
  \alpha\log Z_*=G(\pi^*)
  =\frac12(a^*)^\top Aa^*
  +\frac{\alpha}{2}
  \log\left(\frac{(2\pi\alpha)^m}{\det(A)}\right).
\end{equation}
Applying part~(i) to $f_*$ and any $\pi\in\cP_\pi(\mR^m)$ gives
\begin{equation}
  G(\pi^*)-G(\pi)
  =\alpha\operatorname{KL}(\pi\Vert\pi^*)
  +(\ell^*)^\top\int_{\mR^m}u\pi(u)\,\md u.
\end{equation}
The right-hand side is nonnegative by feasibility and
\eqref{eq:quadratic-kkt}.  Equality forces
$\operatorname{KL}(\pi\Vert\pi^*)=0$, and hence $\pi=\pi^*$ almost
everywhere.  This proves part~(ii).
\end{proof}

\section{Proofs of results in Section~\ref{sec:model}}
\label{app:proofs-model}

\begin{proof}[Proof of Proposition~\ref{prop:qp-geometry}]
Let
\begin{equation}
  F(\mu):=\frac12\mu^\top\Sigma\mu-e^\top\mu.
\end{equation}
Since $\Sigma$ is positive definite, $F$ is coercive and strictly convex.
It therefore has a unique minimizer on the closed convex set $\mR_+^m$.
The first-order optimality condition is
\begin{equation}
  \ell^\top(\mu-\widehat\mu)\ge0,
  \qquad \mu\in\mR_+^m.
\end{equation}
Taking $\mu=\widehat\mu+e_i$ gives $\ell_i\ge0$.  Taking $\mu=0$ and
$\mu=2\widehat\mu$ gives $\ell^\top\widehat\mu=0$.  Conversely, the displayed
conditions imply
$\ell^\top(\mu-\widehat\mu)=\ell^\top\mu\ge0$ for every
$\mu\in\mR_+^m$, so they characterize the minimizer.  Complementarity yields
\begin{equation}
  e^\top\widehat\mu
  =\widehat\mu^\top\Sigma\widehat\mu=\kappa,
\end{equation}
and positive definiteness gives $\kappa\ge0$.

For $a>0$, set $v=a u$.  Then
\begin{equation}
  e^\top u-\frac{a}{2}u^\top\Sigma u
  =\frac1a\left[
  e^\top v-\frac12v^\top\Sigma v\right].
\end{equation}
The expression in brackets has the unique maximizer
$v=\widehat\mu$.  Its value is $\kappa/2$, which proves both scaling
identities \eqref{eq:scailing1} and \eqref{eq:scailing2}.
\end{proof}

\section{Proofs of results in Section~\ref{sec:exploratory}}
\label{app:proofs-exploratory}

\begin{proof}[Proof of Lemma~\ref{lem:crra-partition}]
Set
\begin{equation}
  \Phi_{h,\lambda}(c)
  :=\frac{1}{\lambda}
  \left(\frac{c^{1-\gamma}}{1-\gamma}-hc\right).
\end{equation}
Fix a compact set $K\subset(0,\infty)^2$ of parameters $(h,\lambda)$.
If $0<\gamma<1$, then $c^{1-\gamma}=o(c)$ as $c\to\infty$.
Hence, uniformly over $K$, $\Phi_{h,\lambda}(c)\le-a c$ for all
sufficiently large $c$ and some $a>0$, while $\Phi_{h,\lambda}$ is bounded
above near zero.  If $\gamma>1$, then
\begin{equation}
  \Phi_{h,\lambda}(c)
  =-\frac{1}{\lambda(\gamma-1)c^{\gamma-1}}-\frac{h}{\lambda}c.
\end{equation}
Thus, for some $a_0,a_\infty>0$ independent of $(h,\lambda)\in K$,
\begin{equation}
  \Phi_{h,\lambda}(c)\le-a_0c^{1-\gamma}
  \quad (0<c\le1),
  \qquad
  \Phi_{h,\lambda}(c)\le-a_\infty c
  \quad (c\ge1).
\end{equation}
These bounds show that $Z_\gamma(h,\lambda)$ is finite and positive.  They
also provide $K$-uniform integrable envelopes for
$e^{\Phi_{h,\lambda}(c)}$, $c e^{\Phi_{h,\lambda}(c)}$, and
$c^{1-\gamma}e^{\Phi_{h,\lambda}(c)}$.  Since
$\partial_h e^{\Phi_{h,\lambda}(c)}
=-(c/\lambda)e^{\Phi_{h,\lambda}(c)}$, dominated convergence gives the
claimed continuity properties.  Positivity and continuity imply that
$|\log Z_\gamma|$ is bounded on $K$, so the same envelopes give the stated
uniform first and CRRA moment bounds.  Finally,
\begin{equation}
  |\log\eta_{h,\lambda}(c)|
  \le |\log Z_\gamma(h,\lambda)|
  +\frac{1}{\lambda}\left(
  \frac{c^{1-\gamma}}{|1-\gamma|}+hc\right),
\end{equation}
which proves the claimed uniform bound for the absolute log-density moment.
\end{proof}

\begin{proof}[Proof of Theorem~\ref{thm:crra-ode}]
By \eqref{eq:crra-temperature-range},
$0<\lambda_-\le\lambda_+<\infty$.  Under the reverse-time
change of variables $y(s):=h(T-s)$, equation \eqref{eq:crra-h-ode} becomes
\begin{equation}
  y'(s)=(1-\gamma)\mathfrak G_{\gamma,\lambda}(T-s,y(s)),
  \qquad y(0)=1.
  \label{eq:crra-reverse-ode}
\end{equation}
After extending $\lambda$ as a positive continuous function to a neighborhood
of $[0,T]$, standard ODE theory yields a unique maximal solution and the
comparison, continuation, and Gronwall principles used below
(cf. \citealp[Theorem~1.3, Theorems~2.2 and~2.13,
Corollary~2.15, and Lemma~2.7]{Teschl2012}).

Suppose first that $0<\gamma<1$.  For $0<h\le1$, restricting the integral
defining $Z_\gamma$ to $c\in[1,2]$ gives, uniformly in $t$,
\begin{equation}
  \lambda(t)\log Z_\gamma(h,\lambda(t))\ge-C
\end{equation}
for some constant $C$.  The $h$-independent part of the logarithmic
term is bounded uniformly from below, and hence
\begin{equation}
  \mathfrak G_{\gamma,\lambda}(t,h)
  \ge -C-\frac{m\lambda_-}{2}\log h
  \longrightarrow\infty
  \qquad\text{as }h\downarrow0,
\end{equation}
uniformly in $t$.  Choose $\delta\in(0,1]$ such that
$\mathfrak G_{\gamma,\lambda}(t,\delta)>0$ for all $t$.  Since
$1-\gamma>0$, the constant function $\delta$ is a strict subsolution of the
reverse-time equation \eqref{eq:crra-reverse-ode}.  Scalar comparison yields
$y(s)\ge\delta$ throughout
its maximal interval.

For $h\ge\delta$, the map $h\mapsto Z_\gamma(h,\vartheta)$ is decreasing for
every $\vartheta>0$.  Compactness of $[\lambda_-,\lambda_+]$ therefore gives
\begin{equation}
  \mathfrak G_{\gamma,\lambda}(t,h)\le C(1+h),
  \qquad h\ge\delta,
\end{equation}
with a possibly different $C$.  Thus $y'(s)\le C(1+y(s))$, and Gronwall's
inequality gives $y(s)\le2e^{Cs}-1$ for all $s\ge0$ in the maximal interval.

Now suppose that $\gamma>1$ and assume
\eqref{eq:crra-lower-barrier-condition}.  At the lower boundary,
\begin{equation}
  (1-\gamma)\mathfrak G_{\gamma,\lambda}(t,\underline h)
  \ge0.
\end{equation}
Hence the constant function $\underline h$ is a subsolution of the
reverse-time equation \eqref{eq:crra-reverse-ode}, and scalar comparison gives
$y(s)\ge\underline h$.  For $h\ge\underline h$, restricting the partition
integral to $c\in[1,2]$ gives
\begin{equation}
  \lambda(t)\log Z_\gamma(h,\lambda(t))\ge-C(1+h).
\end{equation}
The Gaussian logarithmic term in \eqref{eq:crra-G-definition} is likewise
bounded below by $-C(1+h)$, so
\begin{equation}
  \mathfrak G_{\gamma,\lambda}(t,h)\ge-C(1+h),
  \qquad h\ge\underline h.
\end{equation}
Since $1-\gamma<0$, this again yields $y'(s)\le C(1+y(s))$ and hence
$y(s)\le2e^{Cs}-1$ for all $s\ge0$ in the maximal interval.  In either case,
$a\le y(s)\le2e^{CT}-1$ for $0\le s\le T$ in the maximal interval, where
$a=\delta$ if $0<\gamma<1$ and $a=\underline h$ if $\gamma>1$.
The solution therefore remains in a fixed compact subset of $(0,\infty)$ up
to any maximal endpoint in $[0,T]$.  The continuation theorem rules
out such an endpoint before $T$.  Translating back to the original time variable
proves the theorem.
\end{proof}

\begin{proof}[Proof of Proposition~\ref{prop:crra-lower-barrier}]
For $h,\vartheta>0$, the change of variables $c=h^{-1/\gamma}y$ gives
\begin{equation}
  Z_\gamma(h,\vartheta)
  =h^{-1/\gamma}\int_0^\infty
  \exp\left[-\frac{h^{(\gamma-1)/\gamma}}{\vartheta}
  \left(y+\frac{1}{(\gamma-1)y^{\gamma-1}}\right)\right]\,\md y.
\end{equation}
The function in parentheses has minimum $\gamma/(\gamma-1)$ at $y=1$ and
is bounded below by $y$.  Splitting the integral at
$y=\gamma/(\gamma-1)$ gives
\begin{equation}
  Z_\gamma(h,\vartheta)
  \le h^{-1/\gamma}
  \exp\left[-\frac{\gamma h^{(\gamma-1)/\gamma}}
  {(\gamma-1)\vartheta}\right]
  \left(
  \frac{\gamma}{\gamma-1}
  +\frac{\vartheta}{h^{(\gamma-1)/\gamma}}
  \right).
  \label{eq:crra-partition-upper-bound}
\end{equation}
The value of $\underline h$ in \eqref{eq:crra-explicit-lower-barrier}
satisfies
\begin{equation}
  \left(r+\frac{\kappa}{2\gamma}\right)\underline h
  \le \frac{\gamma}{2(\gamma-1)}
  \underline h^{(\gamma-1)/\gamma}.
  \label{eq:crra-barrier-linear-bound}
\end{equation}
Applying \eqref{eq:crra-partition-upper-bound} with $h=\underline h$ and
$\vartheta=\lambda(t)$ gives
\begin{align}
  \mathfrak G_{\gamma,\lambda}(t,\underline h)
  \le{}&\left(r+\frac{\kappa}{2\gamma}\right)\underline h
  -\frac{\gamma}{\gamma-1}
  \underline h^{(\gamma-1)/\gamma}\notag\\
  &+\lambda(t)\Bigg[
  \frac m2\log\lambda(t)
  +\frac12\log\left(
  \frac{(2\pi)^m}
  {\gamma^m\underline h^{m+2/\gamma}\det(\Sigma)}
  \right)
  +\log\left(
  \frac{\gamma}{\gamma-1}
  +\frac{\lambda(t)}
  {\underline h^{(\gamma-1)/\gamma}}
  \right)\Bigg].
  \label{eq:crra-barrier-g-bound}
\end{align}
The first bound in \eqref{eq:crra-low-temperature-condition} gives
$\lambda(t)\le\underline h^{(\gamma-1)/\gamma}\le1$.  Hence the expression
in square brackets in \eqref{eq:crra-barrier-g-bound} is at most
\begin{equation}
  \frac12\left|\log\left(
  \frac{(2\pi)^m}
  {\gamma^m\underline h^{m+2/\gamma}\det(\Sigma)}
  \right)\right|
  +\log\left(\frac{2\gamma-1}{\gamma-1}\right).
\end{equation}
The second bound in \eqref{eq:crra-low-temperature-condition} therefore
bounds the $\lambda(t)$-term in \eqref{eq:crra-barrier-g-bound} by
$\gamma\underline h^{(\gamma-1)/\gamma}/[2(\gamma-1)]$.
Combining this estimate with \eqref{eq:crra-barrier-linear-bound} yields
\eqref{eq:crra-lower-barrier-condition}.
\end{proof}

\begin{proof}[Proof of Proposition~\ref{prop:crra-hjb-solution}]
For the ansatz in \eqref{eq:crra-value},
$\partial_xV_\gamma=h(t)x^{-\gamma}$ and
$\partial_{xx}V_\gamma=-\gamma h(t)x^{-\gamma-1}$.
After factoring out $x^{1-\gamma}$, the portfolio maximization in
\eqref{eq:crra-hjb} becomes the problem in Lemma~\ref{lem:quadratic-entropy}(ii)
with
\begin{equation}
  A=\gamma h(t)\Sigma,
  \qquad p=h(t)e,
  \qquad \alpha=\lambda(t).
\end{equation}
The primal minimizer is $\widehat\mu/\gamma$, which gives
\eqref{eq:crra-portfolio}.  The maximized portfolio contribution, after
division by $x^{1-\gamma}$, is
\begin{equation}
  \frac{h(t)\kappa}{2\gamma}
  +\frac{\lambda(t)}{2}
  \log\left(
  \frac{(2\pi\lambda(t))^m}
  {(\gamma h(t))^m\det(\Sigma)}
  \right).
\end{equation}

The consumption maximization reduces to
\begin{equation}
  \sup_{\eta\in\cP_c}
  \left\{
  \int_0^\infty
  \left(\frac{c^{1-\gamma}}{1-\gamma}-h(t)c\right)
  \eta(c)\,\md c+\lambda(t)\cH(\eta)
  \right\}.
\end{equation}
Lemma~\ref{lem:crra-partition} shows that the normalized Gibbs candidate is
well defined and has finite objective and absolute-entropy terms.  The Gibbs
variational formula gives the density in
\eqref{eq:crra-consumption} and the optimal value
$\lambda(t)\log Z_\gamma(h(t),\lambda(t))$.  Substitution of these two
maximized values into \eqref{eq:crra-hjb} yields
\eqref{eq:crra-h-ode}.
\end{proof}

\begin{proof}[Proof of Theorem~\ref{thm:crra-verification}]
Since $h$ and $\lambda$ are continuous and strictly positive on $[0,T]$,
their ranges lie in compact subsets of $(0,\infty)$.  Direct Gaussian
calculation gives
\begin{equation}
  a_{\pi_\gamma^*}=\frac{\widehat\mu}{\gamma},
  \qquad
  M_{\pi_\gamma^*}(s)
  =\frac{\kappa}{\gamma^2}
  +\frac{m\lambda(s)}{\gamma h(s)}.
\end{equation}
The absolute log-density moment of this Gaussian is uniformly bounded in
$s$.  Lemma~\ref{lem:crra-partition}, applied to the compact range of
$(h,\lambda)$, also gives
\begin{equation}
  \sup_{0\le s\le T}\int_0^\infty
  \left(c+c^{1-\gamma}
  +|\log\eta_\gamma^*(c\mid s)|\right)
  \eta_\gamma^*(c\mid s)\,\md c<\infty.
\end{equation}
The explicit densities are jointly Borel measurable.  Their closed-loop
wealth dynamics are a geometric diffusion with bounded deterministic
coefficients and hence have finite positive and negative moments of every
order.  The preceding entropy and consumption bounds then verify all the
conditions in Definition~\ref{def:admissible-policy}.

To verify optimality, let $(\pi,\eta)$ be any admissible policy pair.  Fix
$(s,x)$ and suppress the conditioning variables in its densities.
Recall that $\ell=\Sigma\widehat\mu-e$.
Lemma~\ref{lem:quadratic-entropy} applies to both pointwise functionals.
Apply part~(ii) to the portfolio functional with
\begin{equation}
  A=\gamma h(s)\Sigma,
  \qquad p=h(s)e,
  \qquad \alpha=\lambda(s).
\end{equation}
Its KKT multiplier is $h(s)\ell$.  After division by $x^{1-\gamma}$, the
portfolio and consumption differences are
\begin{align}
  &\int_{\mR^m}\left[h(s)\left(e^\top u-\frac{\gamma}{2}u^\top\Sigma u\right)
    -\lambda(s)\log\pi_\gamma^*(u)\right]\pi_\gamma^*(u)\,\md u\notag\\
  &\quad-\int_{\mR^m}\left[h(s)\left(e^\top u-\frac{\gamma}{2}u^\top\Sigma u\right)
    -\lambda(s)\log\pi(u)\right]\pi(u)\,\md u\notag\\
  &\qquad=\lambda(s)\operatorname{KL}(\pi\Vert\pi_\gamma^*)
    +h(s)\ell^\top a_\pi
  \label{eq:crra-portfolio-gap}
\end{align}
and
\begin{align}
  &\int_0^\infty\left[\frac{c^{1-\gamma}}{1-\gamma}-h(s)c
    -\lambda(s)\log\eta_\gamma^*(c)\right]\eta_\gamma^*(c)\,\md c\notag\\
  &\quad-\int_0^\infty\left[\frac{c^{1-\gamma}}{1-\gamma}-h(s)c
    -\lambda(s)\log\eta(c)\right]\eta(c)\,\md c\notag\\
  &\qquad=\lambda(s)\operatorname{KL}(\eta\Vert\eta_\gamma^*).
  \label{eq:crra-consumption-gap}
\end{align}
Because $\ell\ge0$ and $a_\pi\ge0$, every term on the right-hand sides is
nonnegative.  Combining \eqref{eq:crra-portfolio-gap} and
\eqref{eq:crra-consumption-gap} with the HJB equality, we deduce that
\begin{align}
  &\partial_sV_\gamma(s,x)+\mathcal L^{\pi,\eta}V_\gamma(s,x)
  +\int_0^\infty U_\gamma(xc)\eta(c)\,\md c
  +\lambda(s)x^{1-\gamma}[\cH(\pi)+\cH(\eta)]\notag\\
  &\quad=-x^{1-\gamma}\Big[
  \lambda(s)\operatorname{KL}(\pi\Vert\pi_\gamma^*)
  +h(s)\ell^\top a_\pi
  +\lambda(s)\operatorname{KL}(\eta\Vert\eta_\gamma^*)\Big].
  \label{eq:crra-verification-gap}
\end{align}
Its right-hand side is nonpositive.  For the candidate pair, both relative
entropies vanish and
$h(s)\ell^\top a_{\pi_\gamma^*}=h(s)\ell^\top\widehat\mu/\gamma=0$, so the
residual is zero.
Since $(s,x)$ and the admissible policy pair were arbitrary,
\eqref{eq:crra-verification-gap} holds along the associated closed-loop state
process.

The remaining localization and It\^o verification argument is standard; see,
for example, \citet[Theorem~3.5.2]{Pham2009}.  In the present setting,
Condition~(iv) of Definition~\ref{def:admissible-policy} and the boundedness
of $h$ make the stochastic integral a uniformly integrable martingale, while
Conditions~(ii)--(iii) justify evaluation at $T$.  Applying this argument to
\eqref{eq:crra-verification-gap} and using
$V_\gamma(T,\cdot)=U_\gamma$ gives
$V_\gamma\ge J_\gamma^{\pi,\eta}$ for every admissible pair.  Equality holds
for $(\pi_\gamma^*,\eta_\gamma^*)$, which proves its optimality and identifies
$V_\gamma$ as the value function.
\end{proof}

\begin{proof}[Proof of Proposition~\ref{prop:log-hjb-solution}]
The derivation is the logarithmic counterpart of the proof of
Proposition~\ref{prop:crra-hjb-solution}.  The decomposition
$\log(xc)=\log x+\log c$ separates the wealth dependence, with
$x\partial_xV_{\log}=g(t)$ and
$x^2\partial_{xx}V_{\log}=-g(t)$.  Applying
Lemma~\ref{lem:quadratic-entropy}(ii) to the resulting portfolio maximization
gives \eqref{eq:log-portfolio}.  For consumption, the Gibbs variational
formula applied to $\log c-g(t)c$ gives
\eqref{eq:log-consumption-density}.  Its explicit normalizing constant is
$Z_{\log}(t)$, so the density is the Gamma law in
\eqref{eq:log-consumption}, with maximized value
$\lambda\log Z_{\log}(t)$.  Coefficient matching gives the stated $g$, and
the remaining terms give \eqref{eq:log-k}.  The corresponding
relative-entropy gap identities also give pointwise uniqueness.
\end{proof}

\begin{proof}[Proof of Theorem~\ref{thm:log-verification}]
This is the logarithmic case of
Theorem~\ref{thm:crra-verification}.  The identities
$\log(xc)=\log x+\log c$, $x\partial_xV_{\log}=g$, and
$x^2\partial_{xx}V_{\log}=-g$ remove the wealth-dependent factor from the
CRRA HJB residual.  As $g$ is positive and bounded, the explicit Gaussian
and Gamma policies satisfy the moment and entropy conditions in
Definition~\ref{def:admissible-policy}, and their closed-loop wealth process
is a geometric diffusion with bounded deterministic coefficients.  The
identity $xU'(x)=1$ for $U=\log$ also makes Condition~(iv) straightforward.  The same
entropy-gap and localized It\^o arguments give
$V_{\log}\ge J_{\log}^{\pi,\eta}$ for every admissible pair, with equality for
$(\pi_{\log}^*,\eta_{\log}^*)$.  This proves the
claim.
\end{proof}

\section{Proofs of results in Section~\ref{sec:modal-recovery}}
\label{app:proofs-modal-recovery}

\begin{proof}[Proof of Theorem~\ref{thm:zero-temperature}]
\emph{(i) CRRA value convergence.}
Set
\begin{equation}
  F_\gamma(h):=\frac{\gamma}{1-\gamma}
  h^{(\gamma-1)/\gamma}.
\end{equation}
The function $F_\gamma(h)$ is the maximum over $c>0$ of
$c^{1-\gamma}/(1-\gamma)-hc$, attained at
$c_h:=h^{-1/\gamma}$.  We first record a uniform Laplace estimate for the
partition function.  Define
\begin{equation}
  \chi_\gamma(y)
  :=\frac{\gamma}{1-\gamma}
  -\frac{y^{1-\gamma}}{1-\gamma}+y,
  \qquad y>0.
\end{equation}
This function is nonnegative, has the unique minimum at $y=1$, and satisfies
$\chi_\gamma''(1)=\gamma$.  The change of variables $c=c_hy$ gives
\begin{equation}
  Z_\gamma(h,\lambda)
  =h^{-1/\gamma}\exp\left(\frac{F_\gamma(h)}{\lambda}\right)
  I_\gamma\left(\frac{\lambda}
  {h^{(\gamma-1)/\gamma}}\right),
\end{equation}
where
\begin{align}
  I_\gamma(\tau)
  :=\int_0^\infty
  e^{-\chi_\gamma(y)/\tau}\,\md y,\quad
J_\gamma(\tau):=\int_0^\infty |y-1|
e^{-\chi_\gamma(y)/\tau}\,\md y.
\end{align}
To proceed, we claim that
\begin{align}
  &I_\gamma(\tau)
  =\sqrt{\frac{2\pi\tau}{\gamma}}(1+o(1)),\\
  &\frac{J_\gamma(\tau)}{I_\gamma(\tau)}
  =\sqrt{\frac{2}{\pi\gamma}}\sqrt\tau\,(1+o(1))
  =O(\sqrt\tau)
  \qquad\text{as }\tau\downarrow0.
  \label{eq:zero-temperature-laplace}
\end{align}
Indeed, note that
$\chi_\gamma(y)=\frac{\gamma}{2}(y-1)^2+O(|y-1|^3)$ near $y=1$. 
From the change of variable $y=1+\sqrt\tau z$ in $I_\gamma$ and
$J_\gamma$, it follows that 
\begin{equation}
  \frac{I_\gamma(\tau)}{\sqrt\tau}
  \longrightarrow\int_{\mR}e^{-\gamma z^2/2}\,\md z
  =\sqrt{\frac{2\pi}{\gamma}},
  \qquad
  \frac{J_\gamma(\tau)}{\tau}
  \longrightarrow\int_{\mR}|z|e^{-\gamma z^2/2}\,\md z
  =\frac{2}{\gamma}.
\end{equation}
Here the contributions outside a fixed neighborhood of $1$ are exponentially
smaller because $\chi_\gamma$ is bounded away from zero there and grows at
least linearly as $y\to\infty$.  Taking the ratio proves the second estimate
in \eqref{eq:zero-temperature-laplace}.  When $h$ ranges over a compact
subset of $(0,\infty)$, the quantity $\lambda/h^{(\gamma-1)/\gamma}$ is uniformly
comparable to $\lambda$.  Consequently, it holds by the estimates in
\eqref{eq:zero-temperature-laplace} that
\begin{equation}
  \lambda\log Z_\gamma(h,\lambda)
  =F_\gamma(h)+\frac{\lambda}{2}\log\lambda+O(\lambda)
  \label{eq:zero-temperature-partition-expansion}
\end{equation}
uniformly on every such compact set as $\lambda\downarrow0$.

Define $G_{\gamma,0}(h):=\bigl(r+\kappa/(2\gamma)\bigr)h+F_\gamma(h)$.  Combining
\eqref{eq:zero-temperature-partition-expansion} with the Gaussian term in
\eqref{eq:crra-G-definition}, we get that 
\begin{equation}
  \mathfrak G_{\gamma,\lambda}(t,h)
  =G_{\gamma,0}(h)+\frac{m+1}{2}\lambda(t)\log\lambda(t)
  +O(\lambda(t)),
  \label{eq:zero-temperature-vector-field-expansion}
\end{equation}
locally uniformly in $h$.  Since $\bar\lambda$ is continuous and strictly
positive on $[0,T]$, \eqref{eq:zero-temperature-vector-field-expansion} implies that
for every compact interval $[a,b]\subset(0,\infty)$,
\begin{equation}
  \sup_{\substack{t\in[0,T]\\h\in[a,b]}}
  \left|\mathfrak G_{\gamma,\lambda_\varepsilon}(t,h)
  -G_{\gamma,0}(h)\right|
  \le C_{a,b}\varepsilon|\log\varepsilon|.
  \label{eq:zero-temperature-vector-field-bound}
\end{equation}

To compare the two ODE solutions $h_\varepsilon$ and $h_0$, let
$a_0:=\min_t h_0(t)>0$ and $b_0:=\max_t h_0(t)$, where both extrema are
taken over $[0,T]$.  Set
$y_\varepsilon(s):=h_\varepsilon(T-s)$ and
$y_0(s):=h_0(T-s)$, and let $\tau_\varepsilon$ be the first exit time of
$y_\varepsilon$ from $(a_0/2,2b_0)$, capped at $T$.  Since $G_{\gamma,0}$ is Lipschitz on $[a_0/2,2b_0]$,
on $[0,\tau_\varepsilon]$, we combine
\eqref{eq:zero-temperature-vector-field-bound} with Gronwall's inequality to get
\begin{equation}
  \sup_{0\le u\le s}|y_\varepsilon(u)-y_0(u)|
  \le C\varepsilon|\log\varepsilon|,
  \qquad 0\le s\le\tau_\varepsilon.
  \label{eq:zero-temperature-ode-bound}
\end{equation}
For all sufficiently small $\varepsilon$, the right-hand side is smaller
than the distance from the range of $y_0$ to the boundary of
$[a_0/2,2b_0]$.  Thus the localization never activates,
$\tau_\varepsilon=T$, and
\eqref{eq:zero-temperature-ode-bound} holds on $[0,T]$.
Theorem~\ref{thm:crra-verification} identifies
$V_{\gamma,\varepsilon}$ as the corresponding exploratory value function.
Since $x^{1-\gamma}$ is bounded on each compact subset of $(0,\infty)$,
\eqref{eq:zero-temperature-ode-bound} gives the asserted local value-function
estimate and completes part~(i).

\smallskip
\noindent\emph{(ii) CRRA policy convergence.}
Proposition~\ref{prop:deterministic-solution} gives
$u_\gamma^0(t)=\widehat\mu/\gamma$ and
$c_\gamma^0(t)=h_0(t)^{-1/\gamma}$.  Hence the map
$h\mapsto h^{-1/\gamma}$ is Lipschitz on $[a_0/2,2b_0]$, so
\eqref{eq:zero-temperature-ode-bound} proves the bound for the deployed
consumption mode.

The Gaussian formula in \eqref{eq:crra-portfolio} gives
\begin{equation}
  \sup_{t\in[0,T]}\int_{\mR^m}
  |u-u_\gamma^0(t)|\,
  \pi_{\gamma,\varepsilon}^*(u\mid t,x)\,\md u
  \le\sup_{t\in[0,T]}\left(
  \frac{\lambda_\varepsilon(t)}{\gamma h_\varepsilon(t)}
  \operatorname{tr}(\Sigma^{-1})\right)^{1/2}
  \le C\sqrt\varepsilon.
\end{equation}
To derive the consumption estimate, fix $t\in[0,T]$ and use
$c=h_\varepsilon(t)^{-1/\gamma}y$.  The normalized policy then satisfies
\begin{equation}
  \eta_{\gamma,\varepsilon}^*(c\mid t,x)\,\md c
  =\frac{
  e^{-h_\varepsilon(t)^{(\gamma-1)/\gamma}\chi_\gamma(y)/
  \lambda_\varepsilon(t)}}
  {I_\gamma\!\left(\lambda_\varepsilon(t)/
  h_\varepsilon(t)^{(\gamma-1)/\gamma}\right)}\,\md y.
\end{equation}
The second estimate in \eqref{eq:zero-temperature-laplace} then leads to
\begin{align}
  \int_0^\infty
  |c-h_\varepsilon(t)^{-1/\gamma}|
  \eta_{\gamma,\varepsilon}^*(c\mid t,x)\,\md c
  &=h_\varepsilon(t)^{-1/\gamma}
  \frac{J_\gamma\!\left(\lambda_\varepsilon(t)/
  h_\varepsilon(t)^{(\gamma-1)/\gamma}\right)}
  {I_\gamma\!\left(\lambda_\varepsilon(t)/
  h_\varepsilon(t)^{(\gamma-1)/\gamma}\right)}\notag\\
  &\le C\sqrt{\lambda_\varepsilon(t)}
  h_\varepsilon(t)^{-(\gamma+1)/(2\gamma)}.
\end{align}
By \eqref{eq:zero-temperature-ode-bound}, $h_\varepsilon$ remains in
$[a_0/2,2b_0]$, so the last bound is at most $C\sqrt\varepsilon$ uniformly
in $t$.  Finally, the triangle inequality and the consumption-mode estimate
give
\begin{align}
  \sup_{t\in[0,T]}\int_0^\infty
  |c-c_\gamma^0(t)|\,
  \eta_{\gamma,\varepsilon}^*(c\mid t,x)\,\md c&\le C\sqrt\varepsilon
  +\sup_{t\in[0,T]}
  |h_\varepsilon(t)^{-1/\gamma}-c_\gamma^0(t)|\notag\\
  &\le C\sqrt\varepsilon+C\varepsilon|\log\varepsilon|\\
  &\le C\sqrt\varepsilon.
\end{align}
This proves the remaining consumption estimate and completes part~(ii).

\smallskip
\noindent\emph{(iii) Logarithmic value convergence.}
Theorem~\ref{thm:log-verification} identifies $V_{\log,\varepsilon}$ as the
corresponding exploratory value function.  Write
$V_{\log,\varepsilon}(t,x)=g(t)\log x+k_\varepsilon(t)$ and set
$\Delta k_\varepsilon:=k_\varepsilon-k_0$.  The Stirling formula
\begin{equation}
  \log\Gamma(1+z)
  =\left(z+\frac12\right)\log z-z
  +\frac12\log(2\pi)+R(z),
  \qquad 0<R(z)<\frac{1}{12z},
\end{equation}
applied with $z=1/\varepsilon$ gives, after substitution into
\eqref{eq:log-k},
\begin{align}
  \Delta k_\varepsilon(t)
  ={}&\frac{\varepsilon}{2}\Big\{
  (T-t)\big[(m+1)\log(2\pi\varepsilon)
  -\log\det(\Sigma)\big]\notag\\
  &\quad-(m+2)\big[g(t)\log g(t)-g(t)+1\big]\Big\}
  +(T-t)R_{\log,\varepsilon},
\end{align}
where $R_{\log,\varepsilon}:=\varepsilon R(1/\varepsilon)$ satisfies
$0<R_{\log,\varepsilon}<\varepsilon^2/12$.  The terms multiplying
$\varepsilon$ are uniformly bounded apart from $\log\varepsilon$.  This
proves the global bound in part~(iii).

\smallskip
\noindent\emph{(iv) Logarithmic policy convergence.}
The Gaussian formula in \eqref{eq:log-portfolio} and the identity
$u_{\log}^0(t)=\widehat\mu$ give
\begin{equation}
  \sup_{t\in[0,T]}\int_{\mR^m}
  |u-u_{\log}^0(t)|\,
  \pi_{\log,\varepsilon}^*(u\mid t,x)\,\md u
  \le\sup_{t\in[0,T]}\left(
  \frac{\varepsilon}{g(t)}\operatorname{tr}(\Sigma^{-1})
  \right)^{1/2}
  \le C\sqrt\varepsilon.
\end{equation}
Let $Q_\varepsilon$ have the Gamma distribution with shape
$1+1/\varepsilon$ and rate $1/\varepsilon$.  Then
$\eta_{\log,\varepsilon}^*(\cdot\mid t,x)$ is the law of
$Q_\varepsilon/g(t)$, and
\begin{equation}
  \mE Q_\varepsilon=1+\varepsilon,
  \qquad
  \operatorname{Var}(Q_\varepsilon)=\varepsilon(1+\varepsilon).
\end{equation}
Since $c_{\log}^0(t)=1/g(t)$, the preceding moments imply
\begin{equation}
  \sup_{t\in[0,T]}\int_0^\infty
  |c-c_{\log}^0(t)|\,
  \eta_{\log,\varepsilon}^*(c\mid t,x)\,\md c
  \le\sup_{t\in[0,T]}
  \frac{\sqrt{\varepsilon+2\varepsilon^2}}{g(t)}
  \le\sqrt{3\varepsilon}.
\end{equation}
This proves part~(iv).
\end{proof}

\section{Proofs of results in Section~\ref{sec:policy-improvement}}
\label{app:proofs-policy-improvement}

\begin{proof}[Proof of Theorem~\ref{thm:martingale-characterization}]
For \(\varphi\in\mathcal V\), define the spatial generator associated with
an action pair \((u,c)\) by
\begin{equation}
  \mathcal L^{u,c}\varphi(t,x)
  :=x\left[r+e^\top u-c\right]\partial_x\varphi(t,x)
  +\frac{x^2}{2}u^\top\Sigma u\,\partial_{xx}\varphi(t,x).
\end{equation}
Because the policy pair is scale invariant, Fubini's theorem gives
\begin{equation}
  \mathcal L^{\pi,\eta}\varphi(t,x)
  =\int_{\mR^m}\int_0^\infty
  \mathcal L^{u,c}\varphi(t,x)\nu_t(u,c)\,\md c\,\md u.
  \label{eq:data-averaged-generator}
\end{equation}
Moreover, conditional independence implies
\begin{equation}
  \int_{\mR^m}\int_0^\infty
  \log\nu_t(u,c)\nu_t(u,c)\,\md c\,\md u
  =-\cH(\pi_t)-\cH(\eta_t).
  \label{eq:data-product-entropy}
\end{equation}
Consequently, the fixed-policy running payoff can be written as
\begin{align}
  f_U^{\pi,\eta}(t,x)
  &:={\int_0^\infty U(xc)\eta_t(c)\,\md c}
  +\Lambda(t,x)\left[\cH(\pi_t)+\cH(\eta_t)\right]\notag\\
  &=\int_{\mR^m}\int_0^\infty
  \left[U(xc)-\Lambda(t,x)\log\nu_t(u,c)\right]
  \nu_t(u,c)\,\md c\,\md u.
\end{align}
By the assumed Feynman--Kac characterization, \(J:=J_U^{\pi,\eta}\)
is the unique function in \(\mathcal V\) satisfying
\begin{align}
  &\partial_tJ(t,x)+\mathcal L^{\pi,\eta}J(t,x)
  +f_U^{\pi,\eta}(t,x)=0,
  &&(t,x)\in[0,T)\times(0,\infty),\notag\\
  &J(T,x)=U(x),
  &&x>0.
  \label{eq:data-fixed-policy-fk}
\end{align}

The localized short-time Dynkin formula in
\eqref{eq:little-q-definition} yields
\begin{align}
  q_U^{\pi,\eta}(t,x,u,c)
  ={}&\partial_tJ(t,x)+\mathcal L^{u,c}J(t,x)+U(xc).
  \label{eq:data-q-generator-identity}
\end{align}
Indeed, after applying the localized Dynkin formula, the expression whose
limit defines \(q_U^{\pi,\eta}\) in \eqref{eq:little-q-definition} equals
\begin{equation*}
  \frac{1}{\Delta}\mE_{t,x}^{u,c}
  \int_t^{t+\Delta}
  \left[\partial_sJ(s,X_s^{u,c})
  +\mathcal L^{u,c}J(s,X_s^{u,c})
  +U(X_s^{u,c}c)\right] \,\md s.
\end{equation*}
The asserted identity follows from the assumed integrable limit.

\smallskip
\noindent\emph{Necessity.}
Suppose \((V,q)=(J,q_U^{\pi,\eta})\).  The terminal condition in
\eqref{eq:q-terminal} follows from \eqref{eq:data-fixed-policy-fk}.
Integrating \eqref{eq:data-q-generator-identity} against \(\nu_t\), and
then using \eqref{eq:data-averaged-generator},
\eqref{eq:data-product-entropy}, and \eqref{eq:data-fixed-policy-fk}, gives
\begin{align*}
  &\int_{\mR^m}\int_0^\infty
  \left[q_U^{\pi,\eta}(t,x,u,c)
  -\Lambda(t,x)\log\nu_t(u,c)\right]
  \nu_t(u,c)\,\md c\,\md u\\
  &\quad=\partial_tJ(t,x)+\mathcal L^{\pi,\eta}J(t,x)
  +f_U^{\pi,\eta}(t,x)=0.
\end{align*}
This proves \eqref{eq:q-normalization}.

Fix an initial pair and a finite grid \(\mathcal G\).  By construction,
\(W\) remains an \(\mathbb F^{\mathcal G}\)-Brownian motion, and
\((A_i,C_i)\) is \(\mathcal F_{t_i}^{\mathcal G}\)-measurable.  On
\([t_i,t_{i+1})\), It\^o's formula and
\eqref{eq:data-q-generator-identity} give
\begin{equation}
  \md M_s^{\mathcal G}[J,q_U^{\pi,\eta}]
  =X_s^{\mathcal G}\partial_xJ(s,X_s^{\mathcal G})
  A_i^\top\sigma\,\md W_s.
\end{equation}
Concatenating these identities over the grid and invoking the assumed
uniform integrability proves condition~(iii).

\smallskip
\noindent\emph{Sufficiency.}
Conversely, suppose conditions~(i)--(iii) hold and define
\begin{equation}
  R_{V,q}(t,x,u,c)
  :=\partial_tV(t,x)+\mathcal L^{u,c}V(t,x)+U(xc)-q(t,x,u,c).
\end{equation}
Fix \((t,x)\) with \(t<T\), choose any grid whose first interval is
nondegenerate, and let \((A_0,C_0)\) be the action drawn at \(t\).  On that
first interval, It\^o's formula gives
\begin{align}
  M_s^{\mathcal G}[V,q]-M_t^{\mathcal G}[V,q]
  ={}&\int_t^s
  R_{V,q}(a,X_a^{\mathcal G},A_0,C_0)\,\md a\notag\\
  &+\int_t^s X_a^{\mathcal G}\partial_xV(a,X_a^{\mathcal G})
  A_0^\top\sigma\,\md W_a.
\end{align}
The left-hand side is a uniformly integrable
\(\mathbb F^{\mathcal G}\)-martingale by condition~(iii), and so is the
stochastic integral by assumption.  Their difference is therefore a
continuous finite-variation \(\mathbb F^{\mathcal G}\)-martingale starting
from zero, and hence vanishes indistinguishably.  Since
\(a\mapsto R_{V,q}(a,X_a^{\mathcal G},A_0,C_0)\) is continuous on the first
grid interval, it follows that
\begin{equation*}
  R_{V,q}(t,x,A_0,C_0)=0
  \qquad\text{almost surely}.
\end{equation*}
The law of \((A_0,C_0)\) is \(\nu_t\), so the preceding equality implies
that \(R_{V,q}(t,x,u,c)=0\) for \(\nu_t\)-almost every \((u,c)\).  Since
\(\nu_t\) has full support on \(\mR^m\times(0,\infty)\), the equality
also holds for Lebesgue-a.e. \((u,c)\).  Continuity of
\(R_{V,q}(t,x,\cdot,\cdot)\) then extends the equality to every
\((u,c)\in\mR^m\times(0,\infty)\).  As \((t,x)\) was arbitrary,
\begin{equation}
  q(t,x,u,c)
  =\partial_tV(t,x)+\mathcal L^{u,c}V(t,x)+U(xc)
  \label{eq:data-candidate-q-generator}
\end{equation}
throughout its domain.  Integrating
\eqref{eq:data-candidate-q-generator} against \(\nu_t\) and using
\eqref{eq:q-normalization}, \eqref{eq:data-averaged-generator}, and
\eqref{eq:data-product-entropy}, we obtain
\begin{equation*}
  \partial_tV(t,x)+\mathcal L^{\pi,\eta}V(t,x)
  +f_U^{\pi,\eta}(t,x)=0.
\end{equation*}
Together with \eqref{eq:q-terminal}, this is
\eqref{eq:data-fixed-policy-fk}.  Its assumed uniqueness gives \(V=J\).
Comparing \eqref{eq:data-candidate-q-generator} with
\eqref{eq:data-q-generator-identity} then gives
\(q=q_U^{\pi,\eta}\), completing the proof.
\end{proof}

\begin{proof}[Proof of Theorem~\ref{thm:policy-improvement}]
Suppress the superscript \((\pi,\eta)\) on the current value and
little-\(q\) functions.  For either utility specification, write
\begin{equation*}
  P_{U,t}(u):=\beta_U(t)^\top u-\frac12u^\top K_U(t)u
\end{equation*}
and define
\begin{equation*}
  C_{U,t}(c):=
  \begin{cases}
  \displaystyle\frac{c^{1-\gamma}}{1-\gamma}-h(t)c,
  &U=U_\gamma,\\[0.6em]
  \log c-g(t)c,
  &U=\log.
  \end{cases}
\end{equation*}
The representation is thus
\begin{equation*}
  q_U(t,1,u,c)=d_U(t)+P_{U,t}(u)+C_{U,t}(c).
\end{equation*}
At \(x=1\), the normalization
\eqref{eq:q-normalization} for the current pair is equivalent to
\begin{align}
  0={}&d_U(t)+\int_{\mR^m}P_{U,t}(u)\pi_t(u)\,\md u
  +\Lambda(t,1)\cH(\pi_t)\notag\\
  &+\int_0^\infty C_{U,t}(c)\eta_t(c)\,\md c
  +\Lambda(t,1)\cH(\eta_t).
\end{align}

Set $\ell_U^+(t):=K_U(t)m_U^+(t)-\beta_U(t)$. For the current scale-invariant portfolio policy, write
\begin{equation*}
  a_\pi(t):=\int_{\mR^m}u\pi_t(u)\,\md u\in\mR_+^m.
\end{equation*}
The optimality conditions for the quadratic program in
\eqref{eq:data-portfolio-update} are
\begin{equation*}
  m_U^+(t)\ge0,
  \qquad \ell_U^+(t)\ge0,
  \qquad \ell_U^+(t)^\top m_U^+(t)=0.
\end{equation*}
Applying Lemma~\ref{lem:quadratic-entropy}(ii) with
\(A=K_U(t)\), \(p=\beta_U(t)\), and
\(\alpha=\Lambda(t,1)\) gives the exact portfolio gap
\begin{align}
  &\int_{\mR^m}P_{U,t}(u)\pi_U^+(u\mid t)\,\md u
  +\Lambda(t,1)\cH(\pi_U^+(\cdot\mid t))\notag\\
  &\quad-\int_{\mR^m}P_{U,t}(u)\pi_t(u)\,\md u
  -\Lambda(t,1)\cH(\pi_t)\notag\\
  &=\Lambda(t,1)\operatorname{KL}(\pi_t\Vert\pi_U^+(\cdot\mid t))
  +\ell_U^+(t)^\top a_\pi(t).
  \label{eq:data-portfolio-soft-gap}
\end{align}
The right-hand side is nonnegative as the portfolio mean is
componentwise nonnegative.

The density \(\eta_U^+(\cdot\mid t)\) is the Gibbs density proportional to
\(\exp(C_{U,t}/\Lambda(t,1))\).  For CRRA utility, this follows from
\eqref{eq:data-crra-consumption-update} and the finiteness of the partition
function, while for logarithmic utility, it is exactly the Gamma density in
\eqref{eq:data-log-consumption-update}.  Lemma~\ref{lem:quadratic-entropy}(i)
thus yields
\begin{align}
  &\int_0^\infty C_{U,t}(c)\eta_U^+(c\mid t)\,\md c
  +\Lambda(t,1)\cH(\eta_U^+(\cdot\mid t))\notag\\
  &\quad-\int_0^\infty C_{U,t}(c)\eta_t(c)\,\md c
  -\Lambda(t,1)\cH(\eta_t)
  =\Lambda(t,1)\operatorname{KL}(\eta_t\Vert\eta_U^+(\cdot\mid t)).
  \label{eq:data-consumption-soft-gap}
\end{align}
Admissibility of the current and updated policy pairs makes all reward and
entropy terms on the left-hand sides of
\eqref{eq:data-portfolio-soft-gap} and
\eqref{eq:data-consumption-soft-gap} finite.  For the portfolio terms,
positive definiteness of \(K_U(t)\) and \(\Sigma\) makes their quadratic
forms equivalent at each fixed \(t\).  The two gap identities then imply
that the displayed KL divergences are finite.

Let \(\nu_t^+(u,c):=\pi_U^+(u\mid t)\eta_U^+(c\mid t)\).  Define
\begin{align}
  \mathcal R_U^+(t,x)
  &:={\int_{\mR^m}\int_0^\infty
  \left[q_U(t,x,u,c)-\Lambda(t,x)\log\nu_t^+(u,c)\right]
  \nu_t^+(u,c)\,\md c\,\md u}.
\end{align}
Adding \eqref{eq:data-portfolio-soft-gap} and
\eqref{eq:data-consumption-soft-gap} and using the current policy
normalization \eqref{eq:q-normalization} at \(x=1\) gives
\begin{align}
  \mathcal R_U^+(t,1)
  ={}&\Lambda(t,1)\bigl[\operatorname{KL}(\pi_t\Vert\pi_U^+(\cdot\mid t))
  +\operatorname{KL}(\eta_t\Vert\eta_U^+(\cdot\mid t))\bigr]\notag\\
  &+\ell_U^+(t)^\top a_\pi(t)\ge0.
\end{align}
For CRRA utility, \(q_U\) and \(\Lambda\) both scale with \(x^{1-\gamma}\),
so \(\mathcal R_U^+(t,x)=x^{1-\gamma}\mathcal R_U^+(t,1)\). For logarithmic
utility, both are independent of \(x\), so
\(\mathcal R_U^+(t,x)=\mathcal R_U^+(t,1)\). Thus
\(\mathcal R_U^+(t,x)\ge0\) for every \(x>0\).
By Theorem~\ref{thm:martingale-characterization} and
\eqref{eq:data-q-generator-identity}, we have
\begin{align}
  \mathcal R_U^+(t,x)
  ={}&\partial_tJ_U^{\pi,\eta}(t,x)
  +\mathcal L^{\pi_U^+,\eta_U^+}J_U^{\pi,\eta}(t,x)
  +\int_0^\infty U(xc)\eta_U^+(c\mid t)\,\md c\notag\\
  &+\Lambda(t,x)\left[
  \cH(\pi_U^+(\cdot\mid t))+\cH(\eta_U^+(\cdot\mid t))
  \right].
\end{align}

Let \(X^+\) denote the relaxed wealth process under the updated policy pair.
Applying the assumed Dynkin formula to the current value function along
\(X^+\), using \(J_U^{\pi,\eta}(T,\cdot)=U\), and comparing with the
fixed-policy objective gives
\begin{align*}
  J_U^{\pi_U^+,\eta_U^+}(t,x)-J_U^{\pi,\eta}(t,x)
  =\mE_{t,x}^{\pi_U^+,\eta_U^+}
  \int_t^T\mathcal R_U^+(s,X_s^+)\,\md s\ge0.
\end{align*}
This proves the assertion for \(t<T\).  At \(t=T\), both values equal
\(U(x)\), so the result holds on the stated domain.
\end{proof}

\begin{proof}[Proof of Theorem~\ref{thm:exact-policy-iteration-convergence}]
Write $J_n=J_U^{\pi_n,\eta_n}$. Under scale invariance,
the law of $X_s/x$ is independent of $x>0$. For a fixed action
$(u,c)$ this ratio is lognormal, so It\^o's formula in
\eqref{eq:little-q-definition} gives
$q_n=\partial_tJ_n+\mathcal L^{u,c}J_n+U(xc)$ for each iterate $J_n$ considered below.

\smallskip
\noindent\emph{(i) Logarithmic utility.}
By \eqref{eq:log-objective}, $J_n(t,x)=g(t)\log x+J_n(t,1)$ with
$g(t)=1+T-t$. The action-dependent part of $q_n$ is
\begin{equation}
 g(t)\bigl(e^\top u-\tfrac12 u^\top\Sigma u\bigr)+\log c-g(t)c.
\end{equation}
Thus every update gives \eqref{eq:log-portfolio}--\eqref{eq:log-consumption},
with modes $\widehat\mu$ and $1/g(t)$.

\smallskip
\noindent\emph{(ii) CRRA policy densities.}
For simplicity, set $G=\mathfrak G_{\gamma,\lambda}$, and let $h^*>0$ solve
\eqref{eq:crra-h-ode} on $[0,T]$. By \eqref{eq:crra-objective},
$J_n(t,x)=h_n(t)x^{1-\gamma}/(1-\gamma)$, hence $h_n(t)=(1-\gamma)J_n(t,1)$. Define
\begin{equation}
 F_n(t,z)=z\left(r+e^\top a_{\pi_n}-\bar c_{\eta_n}
                         -\tfrac{\gamma}{2}M_{\pi_n}\right)
   +\frac{1}{1-\gamma}\int_0^\infty c^{1-\gamma}\eta_n(c\mid t)\,\md c
   +\lambda(t)\bigl[\cH(\pi_n)+\cH(\eta_n)\bigr].
\end{equation}
Fixed-policy evaluation gives $h_n'/(1-\gamma)+F_n(t,h_n)=0$, $h_n(T)=1$.
By the pointwise maximization in Proposition~\ref{prop:crra-hjb-solution},
$F_n(t,z)\le G(t,z)$ for $z>0$. At $z=h_n(t)>0$ the portfolio
coefficients are $(h_ne,\gamma h_n\Sigma)$, and the maximizing update is
\begin{equation}
 \pi_{n+1}(\cdot\mid t)=\N\left(\frac{\widehat\mu}{\gamma},
             \frac{\lambda(t)}{\gamma h_n(t)}\Sigma^{-1}\right),
 \qquad \eta_{n+1}(\cdot\mid t)=\eta_{h_n(t),\lambda(t)}.
 \label{eq:pi-crra-exact-update}
\end{equation}
For positive $h_n\in C^1([0,T])$, the parameters $(h_n,\lambda)$ have
compact range in $(0,\infty)^2$. The moment bounds in the proof of
Theorem~\ref{thm:crra-verification} then give admissibility of this update.
The policy pair in \eqref{eq:pi-crra-exact-update} attains the maximum
defining $G(t,h_n(t))$, so $F_{n+1}(t,h_n(t))=G(t,h_n(t))$.
Since $F_{n+1}(t,z)\le G(t,z)$ for every $z>0$, the differentiable function
$z\mapsto G(t,z)-F_{n+1}(t,z)$ has an interior minimum at $z=h_n(t)$.
Hence $\partial_zF_{n+1}(t,h_n(t))=\partial_zG(t,h_n(t))$.
As $F_{n+1}(t,\cdot)$ is affine, these equalities give
\begin{align}
 &F_{n+1}(t,z)=G(t,h_n(t))+\partial_zG(t,h_n(t))(z-h_n(t)).
 \label{eq:pi-crra-newton-recursion}
\end{align}

Put $k_n(t,s)=\exp\bigl((1-\gamma)\int_t^s \partial_zG(v,h_n(v))\,\md v\bigr)$. Subtracting the equations
of $h_n$, $h_{n+1}$ (which are linear) and $h^*$, all with terminal value one, gives
\begin{align}
 \frac{h_{n+1}(t)-h_n(t)}{1-\gamma}
 &=\int_t^T k_n(t,s)\bigl[G(s,h_n(s))-F_n(s,h_n(s))\bigr]\,\md s\ge0,\notag\\
 \frac{h^*(t)-h_{n+1}(t)}{1-\gamma}
 &=\int_t^T k_n(t,s)\bigl[G(s,h^*(s))-F_{n+1}(s,h^*(s))\bigr]\,\md s\ge0.
 \label{eq:pi-crra-error-representation}
\end{align}
Thus $h_n(t)\leq h_{n+1}(t)\leq h^*(t)$ for either sign
of $1-\gamma$. Induction from $h_0>0$ puts all $h_n,h^*$ in a common interval
$[a,b]\subset(0,\infty)$.

The Gibbs kernel bounds in the proof of Lemma~\ref{lem:crra-partition}
remain integrable after multiplication by $c^2$. Differentiating
\eqref{eq:crra-G-definition} on $[0,T]\times[a,b]$ gives
\begin{equation}
 0\le\partial_{zz}G(t,z)=\frac{m\lambda(t)}{2z^2}
 +\frac{\operatorname{Var}_{\eta_{z,\lambda(t)}}(c)}{\lambda(t)}
 \le L<\infty.
\end{equation}
Set $d_n=h_n-h^*$ and $E_n=\|d_n\|_\infty$. By
\eqref{eq:pi-crra-newton-recursion} and Taylor's theorem,
\begin{equation}\label{eq:taylor-bound}
 0\le G(t,h^*(t))-F_{n+1}(t,h^*(t))\le \tfrac{L}{2}|d_n(t)|^2.
\end{equation}
Since $\partial_zG$
is bounded on the same compact set, \eqref{eq:pi-crra-error-representation}
yields, for some $B>0$ independent of $n$,
\begin{equation}
 |d_{n+1}(t)|\le B\int_t^T|d_n(s)|^2\,\md s,\qquad E_{n+1}\le BT E_n^2.
\end{equation}
Using $|d_n|\le b-a$ and iterating the integral gives
$E_n\le E_0[BT(b-a)]^n/n!$. Choose $N$ with $BT E_N\le1/2$. Then
$BT E_{N+j}\le2^{-2^j}$ for $j\ge0$, hence
\begin{equation}
 E_n\le C_h\exp(-\alpha 2^n),\qquad \alpha=2^{-N}\log2,\quad n\ge0,
 \label{eq:pi-crra-coefficient-bound}
\end{equation}
with $C_h$ enlarged to cover $n<N$.

Since $\ell^\top a_{\pi_{n+1}}=\ell^\top\widehat\mu/\gamma=0$,
\eqref{eq:crra-portfolio-gap} and \eqref{eq:crra-consumption-gap} give
\begin{equation}
 G(t,h^*(t))-F_{n+1}(t,h^*(t))
 =\lambda(t)\bigl[\operatorname{KL}(\pi_{n+1}\Vert\pi_\gamma^*)
                 +\operatorname{KL}(\eta_{n+1}\Vert\eta_\gamma^*)\bigr].
\end{equation}
Pinsker's inequality, Cauchy--Schwarz and \eqref{eq:taylor-bound} yield
\begin{equation}
 \|\pi_{n+1}-\pi_\gamma^*\|_1+\|\eta_{n+1}-\eta_\gamma^*\|_1
 \le\sqrt{2L/\lambda_-}\,E_n.
\end{equation}
Now \eqref{eq:pi-crra-coefficient-bound} gives
\eqref{eq:pi-crra-density-convergence}, including $t=T$ by continuity.

\smallskip
\noindent\emph{(iii) CRRA policy modes.}
For $n\ge1$, the Gaussian in \eqref{eq:pi-crra-exact-update} has mode
$\widehat\mu/\gamma$. Since $\partial_c\log\eta_n(c\mid t)
=(c^{-\gamma}-h_{n-1}(t))/\lambda(t)$, the consumption modes are
$m_n=h_{n-1}^{-1/\gamma}$ and $m^*=(h^*)^{-1/\gamma}$.
The mean value theorem on $[a,b]$ gives
\begin{equation}
 \|m_n-m^*\|_\infty\le\frac{a^{-1-1/\gamma}}{\gamma}E_{n-1}.
\end{equation}
By \eqref{eq:pi-crra-coefficient-bound}, both
\eqref{eq:pi-crra-density-convergence} and \eqref{eq:pi-crra-mode-convergence}
hold with $C=C_h\max\{\sqrt{2L/\lambda_-},a^{-1-1/\gamma}/\gamma\}$.
\end{proof}

\end{document}